\documentclass[a4paper]{quantumarticle}
\pdfoutput=1
\usepackage[utf8]{inputenc}
\usepackage[english]{babel}
\usepackage[T1]{fontenc}
\usepackage{amsmath}
\usepackage{amssymb}
\usepackage{hyperref}
\usepackage{orcidlink}

\usepackage{tikz}
\usepackage{lipsum}

\begin{document}

\title{MAQCY: Modular Atom-Array Quantum Computing with Space-Time Hybrid Multiplexing}

\author{Andrew Byun\,\orcidlink{0000-0001-8306-1818}}
\thanks{These two authors contributed equally.}
\affiliation{Department of Physics, Korea University, 145 Anam-ro, Seongbuk-gu, Seoul 02841, Republic of Korea}
\altaffiliation{Present address: Institute for Theoretical Physics, University of Innsbruck, A-6020 Innsbruck, Austria}
\orcid{0000-0001-8306-1818}

\author{Chanseul Lee}
\thanks{These two authors contributed equally.}
\affiliation{Department of Physics, Korea University, 145 Anam-ro, Seongbuk-gu, Seoul 02841, Republic of Korea}

\author{Eunsik Yoon}
\affiliation{Department of Physics, Korea University, 145 Anam-ro, Seongbuk-gu, Seoul 02841, Republic of Korea}

\author{Minhyuk Kim\,\orcidlink{0000-0001-8705-7795}}
\email[]{minhyukkim@korea.ac.kr}
\affiliation{Department of Physics, Korea University, 145 Anam-ro, Seongbuk-gu, Seoul 02841, Republic of Korea}

\author{Tai Hyun Yoon\,\orcidlink{0000-0002-2408-9295}}
\email[]{thyoon@korea.ac.kr}
\affiliation{Department of Physics, Korea University, 145 Anam-ro, Seongbuk-gu, Seoul 02841, Republic of Korea}

\maketitle

\begin{abstract}
 We present a modular atom-array quantum computing architecture with space-time hybrid multiplexing (MAQCY), a dynamic optical tweezer-based protocol for fully connected and scalable universal quantum computation. By extending the concept of globally controlled static dual-species Rydberg atom wires~\cite{Cesa23}, we develop an entirely new approach using Q-Pairs, which consist of globally controlled and temporally multiplexed dual-species Rydberg blockaded atom and superatom pairs. Space-time hybrid multiplexing of Q-Pairs achieves $\mathcal{O}(N)$ linear scaling in the number of required physical qubits, while preserving coherence and mitigating circuit-depth limitations through \textit{in-situ} atom replacement. To demonstrate MAQCY’s versatility, we implement a three-qubit quantum Fourier transform using only global operations and atom transport. We also propose a concrete implementation using ytterbium isotopes, paving the way toward large-scale, fault-tolerant quantum computing.
\end{abstract}

\section{INTRODUCTION}
Neutral atom arrays, formed by individually trapped atoms, have become a key platform for quantum science and technology~\cite{Saffman10, Browaeys2020, Kim23}. 
These platforms have rapidly advanced over the past decade, demonstrating remarkable experimental progress due to their relatively versatile connectivity and scalability compared to other architectures.

In particular, \textit{Rydberg blockade}~\cite{Jaksch00, Lukin01, Urban09, Gaetan09} and \textit{moving tweezers}~\cite{Kim16, Barredo16, Endres16, Bluvstein22, Bluvstein23, Shaw2024} represent key milestones in neutral-atom-based quantum computing.
The Rydberg blockade mediates entanglement by leveraging strong interactions between Rydberg atoms to suppress multiple excitations within a certain volume~\cite{Wilk10, Isenhower10, Levine19}. 
Moving tweezers have been used to generate scalable, defect-free arrays by filling vacancies arising from stochastic atom loading~\cite{Kim16, Barredo16, Endres16, Norcia2024, Pichard2024}. 
Moreover, since atoms can be transported without destroying coherence~\cite{Bluvstein22}, this technique has been extended to realize all-to-all connectivity between atoms$-$an essential requirement for universal quantum computing.
Exploiting these features, quantum platforms based on neutral atom arrays have made significant advances in analog~\cite{Ebadi2022, Kim2022, Graham2022} and digital quantum computation~\cite{Bluvstein23, Reichardt2024}, many-body physics~\cite{Browaeys2020, Bernien2017, Semeghini2021, Bluvstein2021}, and quantum metrology~\cite{Shaw2024}.

Following the development of alkali atomic arrays, dual-alkali species arrays$-$offering versatile interaction channels via Rydberg states~\cite{Beterov2015, Zeng2017, Singh2022, Anand2024, Singh2023, Sheng2022}$-$and arrays of alkaline-earth-like atoms (AEAs)~
\cite{Norcia2018, Cooper2018, Saskin2019, Madjarov2020, tsai2024} with two valence electrons have recently emerged. 
In particular, AEAs are distinguished by a narrow-linewidth clock transition between spin-singlet ground and spin-triplet excited states, both with zero total electronic angular momentum, enabling intrinsically long coherence times and high-fidelity operations at optical frequencies. By exploiting multiple electronic levels, both optical clock qubits~\cite{Norcia2018, Cooper2018, Saskin2019} and nuclear spin qubits in fermionic isotopes~\cite{Jenkins2022, Ma2022, Barnes2022} have been realized, each offering favorable intrinsic qubit coherence. Furthermore, mid-circuit measurement~\cite{Norcia2023, Lis2023, Huie2023} combined with erasure operations~\cite{Wu2022, Scholl2023, Ma2023} provides effective decoherence mitigation. Recently, various quantum platforms based on AEAs have demonstrated both the Rydberg blockade~\cite{Madjarov2020, tsai2024} and moving tweezers~\cite{Shaw2024, Norcia2024}, showcasing their favorable properties for near-term quantum platforms.

For controlling qubits in the array, local addressing~\cite{Urban09, Bluvstein23, Isenhower10, Graham2022, Omran2019} was first demonstrated in alkali atomic systems.
While local addressing has proven successful, its implementation remains challenging and has so far been limited to restricted geometries~\cite{Graham2022, Omran2019, Chen2023, deoliveira2024}. To circumvent the need for tightly focused multiple local addressing beams, Cesa and Pichler (CP) recently proposed a universal quantum computing protocol based solely on a globally driven laser beam. Their scheme employs a two-dimensional \textit{static} arrangement of dual-species atoms, where the spatial configuration is designed to match a given quantum circuit. Within this architecture, superatoms~\cite{Dudin2012, Ebert2014, Ebert2015, Zeiher2015, Labuhn2016}, collectively acting atomic clusters, combined with composite pulses, enable effective local control using only a globally driven laser beam. However, achieving full connectivity requires a large physical qubit overhead, which in turn may increase susceptibility to vacuum-limited survival times of atoms trapped in optical tweezers.

\begin{figure*}[]
	\includegraphics[width=1.0\textwidth]{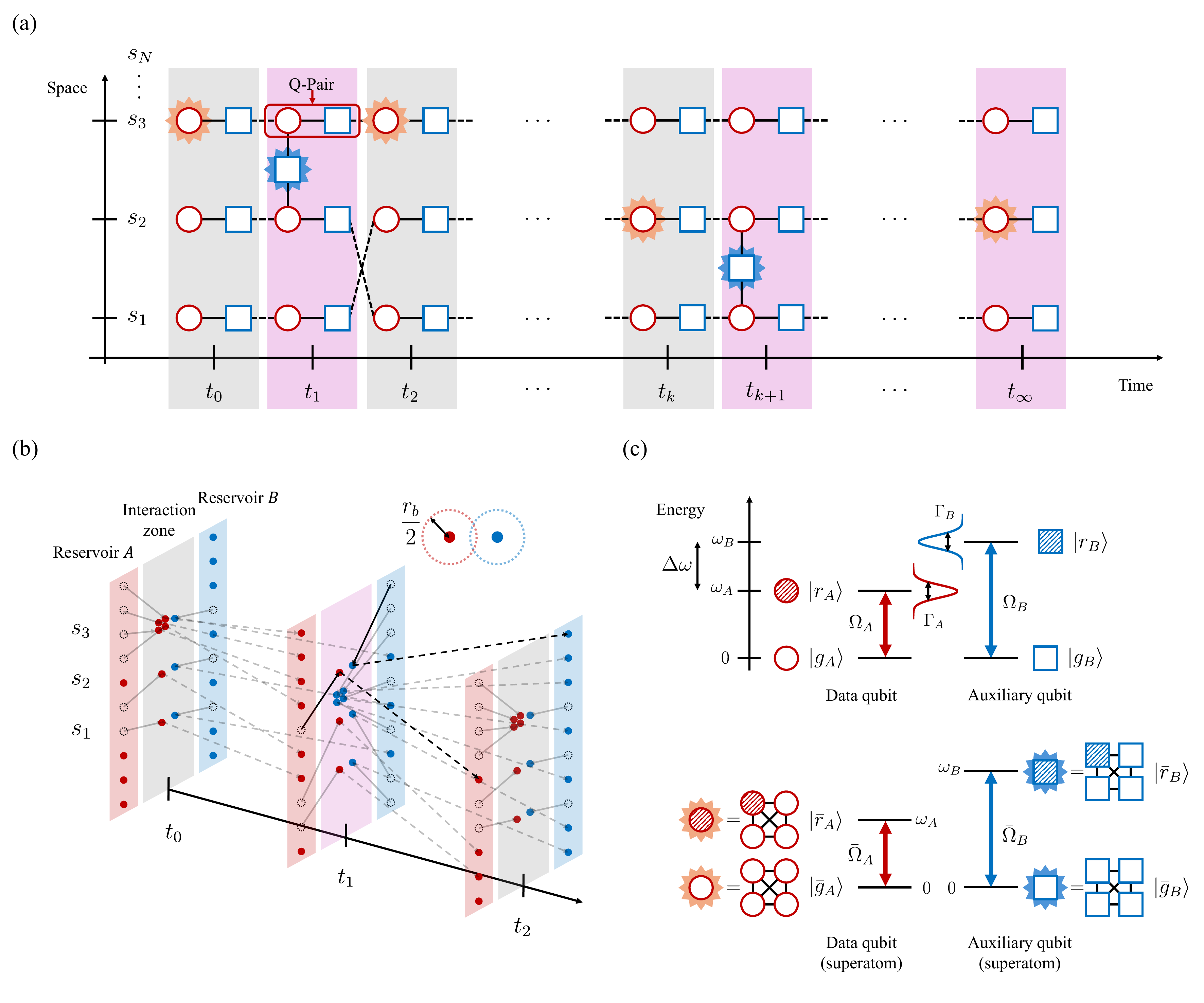}
\caption{
\textbf{MAQCY; Modular atom-array quantum computing protocol with space-time hybrid multiplexing.} 
\textbf{(a)} Space-time hybrid array of Q-Pairs. A Q-Pair consists of a dual-species atomic pair: a red circle (data qubit, species $A$) and a blue square (auxiliary qubit, species $B$). Local control is realized using a superatom$-$an ensemble of atoms$-$indicated by a star-shaped background. Each Q-Pair is encoded in a hybrid mode $(t_k, s_l)$ as $\left|\Psi(t_k, s_l)\right\rangle$.
\textbf{(b)} Zoned architecture for experimental realization. All Q-Pairs are arranged in the interaction zone, where a driving laser couples ground and Rydberg states. The laser is applied only within this zone. Rydberg blockade mediates correlations within a blockade volume (indicated by a dotted circle). During temporal mode translation, $\tilde{\mathcal{T}}: t_k \rightarrow t_{k+1}$, atoms from the reservoirs replace those in the interaction zone.
\textbf{(c)} Energy-level diagrams of the dual-species AEAs. The two atomic species are spectrally distinguishable, allowing global laser beams to address individual atoms in each Q-Pair. A superatom (e.g., with $N=4$ atoms) functions as an effective data qubit and serves as a local (target) data qubit in the MAQCY protocol. Additional details are provided in the main text.}
\label{Fig1}
\end{figure*}

We propose MAQCY, a universal quantum computing protocol for a neutral-atom platform that utilizes a modular array with space-time hybrid multiplexing. The core of MAQCY consists of two key elements: (1) Q-Pairs and (2) space-time hybrid multiplexing. Q-Pairs are a new building block of MAQCY: a Rydberg-blockaded pair composed of atoms or superatoms. The Q-Pair concept is inspired by the quantum wire in the CP protocol~\cite{Cesa23}, which acts as the basic carrier of quantum information.
This Q-Pair provides more favorable scaling than the original wire-based architecture.

Space-time hybrid multiplexing connects two Q-Pairs that are separated in both space and time. By dynamically loading and discarding (super)atoms with moving optical tweezers, analogous to temporal multiplexing~\cite{Larsen19, Asavanant19, Larsen21} in measurement-based quantum computing~\cite{Briegel2009}, space-time hybrid multiplexing enables Rydberg-blockade-mediated information flow between temporally distinct Q-Pairs.

In this paper, we show that the MAQCY protocol can be realized on state-of-the-art experimental platforms based on dual isotopes of ytterbium atoms~\cite{Reichardt2024, Ma2022, Huie2023, Ma2023}. Moreover, it can also be realized using other AEA species possessing similar atomic structures.

\paragraph{Q-Pairs:}
Figure~\ref{Fig1}(a) illustrates an example of quantum circuit implementation using the MAQCY protocol. We begin by constructing a Q-Pair (rounded red box) within a single temporal mode. Each Q-Pair comprises dual-species atoms: atom $A$ (red circle) serves as the \textit{data qubit}, storing quantum information, and atom $B$ (blue square) always acts as the single \textit{auxiliary qubit}, facilitating quantum information flow and entanglement.

Superatoms (star-shaped background) are utilized to implement single- and two-qubit gates. The data qubit of a Q-Pair can be either a single atom or a superatom. By contrast, the auxiliary qubit of a Q-Pair should always be a single atom of species $B$. Single-qubit gates are performed on superatom $A$ by applying global pulse sequences resonant with species $A$ only, leaving all other atoms unaffected. Two-qubit entanglement between adjacent Q-Pairs is achieved by placing superatom $B$ to mediate correlations between data qubits.

\paragraph{Space-time hybrid multiplexing:}
The arrangement of Q-Pairs is mapped onto a space-time hybrid mode $(t_k, s_l)$, as shown in Fig.~\ref{Fig1}(a), where $0 \le k < \infty$ and $1 \le l \le N$. Here, the temporal mode $t_k$ denotes the stage of circuit operation, visually indicated by alternating background colors: gray for even $k$ and pink for odd $k$. 
The spatial mode $s_l$ represents the spatial position of a Q-Pair. In our MAQCY protocol, each temporal mode is translated by a 
temporal mode translation operator $\tilde{\mathcal{T}}$ (see Eq.~(\ref{EqT})). Similarly, the spatial positions of two Q-Pairs located at $s_l$ and $s_m$ can be interchanged by a SWAP gate $\tilde{S}(s_l, s_m)$ (see Eq.~(\ref{Swap})). Therefore, the quantum state $\left|\Psi(t_k, s_l)\right\rangle$ of Q-Pairs can be coherently transformed (teleported) by two unitary operators $\tilde{\mathcal{T}}$ and $\tilde{\mathcal{S}}^{(s_l,s_m)}$ acting on the temporal and spatial degrees of freedom, respectively:
\begin{subequations}
\begin{eqnarray}
&&\left|\Psi(t_{k+1}, s_l)\right\rangle = \tilde{\mathcal{T}} \left|\Psi(t_k, s_l)\right\rangle, \label{Eq_TempMux} \\
&&\left|\Psi(t_k, s_m)\right\rangle \left|\Psi(t_k, s_l)\right\rangle = \nonumber \\ && \qquad  \qquad \tilde{\mathcal{S}}^{(s_l,s_m)} \left|\Psi(t_k, s_l)\right\rangle \left|\Psi(t_k, s_m)\right\rangle.\label{Eq_SparMux}
\end{eqnarray}
\end{subequations}
Note that the temporal mode translation $\tilde{\mathcal{T}}$ generates correlations between two temporally neighboring Q-Pairs that remain in the same spatial mode in the absence of a SWAP $\tilde{\mathcal{S}}$. By analogy with temporal multiplexing in photonic systems~\cite{Larsen19, Asavanant19, Larsen21}, we refer to this mode translation as space-time multiplexing.

Figure~\ref{Fig1}(b) depicts a plausible \textit{zoned-architecture}-based~\cite{Bluvstein22, Bluvstein23, Reichardt2024} implementation of MAQCY. Both species $A$ and $B$ are initially prepared in the reservoir. All atoms comprising the Q-Pairs are shuttled into an interaction zone, where a globally driven laser is applied. In these Q-Pairs, the data and auxiliary atoms are positioned closer than their respective Rydberg blockade radii $r_b$ (indicated by red and blue dotted circles), enabling strong interactions. During temporal mode translation, atoms from the reservoir replace or renew the Q-Pair atoms. 
The black arrows in Fig.~\ref{Fig1}(b) indicate exemplary atomic trajectories used to construct the Q-Pair in Fig.~\ref{Fig1}(a). This dynamic qubit rearrangement enhances both scalability and keeps the quantum coherence across replaced atoms.

\section{MAQCY protocol}
\subsection{Rydberg atom system}
The qubit states of atoms of species $A$ and $B$ are coupled independently via chromatically distinct resonant fields~\cite{Zeng2017, Singh2023, Anand2024}, with transition frequency $\omega_{\mu}$ and Rabi frequency $\Omega_{\mu}$, respectively. These couplings satisfy the condition $\Delta\omega = |\omega_B - \omega_A| \gg \Omega_\mu \gg \Gamma_\mu$, where $\Gamma_\mu$ is the decay rate, as illustrated in Fig.~\ref{Fig1}(c).

The Hamiltonian of the Rydberg atom system, $\mathcal{H} = \mathcal{H}_{\rm dri} + \mathcal{H}_{\rm int}$, is given by~\cite{Cesa23}
\begin{subequations}\label{E1}
\begin{eqnarray}
\mathcal{H}_{\rm dri} &=& \frac{\hbar}{2}\sum_{\mu} \Omega_{\mu}  e^{i\phi_{\mu}}\left|g_{\mu}\right\rangle\left\langle r_{\mu}\right| + \text{h.c.}, \label{E1a}\\
\mathcal{H}_{\rm int} &=& U_{\rm ho} + V_{\rm he}, \label{E1b}
\end{eqnarray}
\end{subequations}
where $\phi_{\mu}$ is the phase of the driving field for species $\mu$.

There are two types of Rydberg interaction energies: $U_{\rm ho}$ between atoms of the same species, and $V_{\rm he}$ between atoms of different species,
\begin{subequations}\label{E2}
\begin{eqnarray}
U_{\rm ho} &=&  \sum_{\mu} \sum_{i\neq j} U_{\mu} (r_{i,j}) \left|r_\mu r_\mu\right\rangle\left\langle r_\mu r_\mu\right|, \label{E2a}\\
V_{\rm he} &=&  \sum_{\mu\neq\nu} \sum_{i\neq j} V (r_{i,j}) \left|r_\mu r_\nu\right\rangle\left\langle r_\mu r_\nu\right|,\label{E2b}
\end{eqnarray}
\end{subequations}
where $r_{i,j}$ is the distance between atoms $i$ and $j$. Here, $U_\mu$ denotes the Rydberg interaction energy between homogeneous atoms, and $V$ that between heterogeneous atoms~\cite{Beterov2015, Anand2024}.

If two atoms, regardless of species, are placed close enough such that $U_\mu(r_{i,j}), V(r_{i,j}) \gg \Omega_\mu$, the Rydberg blockade becomes active. We assume the PXP model~\cite{Lesanovsky2012, Serbyn2021}, where interactions outside the blockade radius $r_b$ are ignored.

\subsection{Superatom}
A superatom is a correlated atomic system composed of $M \geq 2$ atoms, coupled via Rydberg blockade, with an enhanced collective Rabi frequency $\bar{\Omega}_\mu = \sqrt{M}\Omega_\mu$~\cite{Dudin2012, Ebert2014, Ebert2015, Zeiher2015, Labuhn2016}. It functions as an effective qubit with two basis states, $\left|\bar{g}_\mu\right\rangle$ and $\left|\bar{r}_\mu\right\rangle$, defined as (see Fig.~\ref{Fig1}(c)):
\begin{subequations}
\begin{eqnarray}
\left|\bar{g}_\mu\right\rangle &=& \left|g_\mu g_\mu \cdots g_\mu\right\rangle, \\
\left|\bar{r}_\mu\right\rangle &=& \frac{\left|r_\mu g_\mu \cdots g_\mu \right\rangle + \cdots + \left|g_\mu g_\mu \cdots r_\mu\right\rangle}{\sqrt{M}}.
\end{eqnarray}
\end{subequations}

Examples of superatoms $A$ and $B$ with $M = 4$ atoms each (yielding $\bar{\Omega}_\mu = 2\Omega_\mu$) are shown in Fig.~\ref{Fig1}(c). Due to the collective enhancement, superatoms respond differently to global driving fields than single atoms. This differential behavior allows for local controllability in MAQCY, as discussed in the CP protocol~\cite{Cesa23}: quantum operations act only on Q‑Pairs with superatomic data qubits, while single-atom qubits remain unaffected.

All quantum operations in the MAQCY protocol are classified based on whether they act on single atoms, superatoms, or both. The corresponding operator notations are summarized in Table~\ref{tab:table0}.\\

\begin{table*}[t]
\caption{\label{tab:table0}%
Summary of gate operator notations used in the MAQCY protocol.}
\centering
\renewcommand{\arraystretch}{1.3}
\begin{tabular}{p{3cm} p{13cm}}
\hline\hline
\textbf{Notation} & \textbf{Description} \\
\hline
$\hat{G}$ & Operator acting only on single-atom qubits. \\
$\bar{G}$ & Operator acting only on superatomic qubits. \\
$\bar{\bar{\mathcal{G}}} = \begin{cases}
\hat{\mathbb{I}} \\
\bar{G}
\end{cases}$ & Global operator that selectively acts on superatoms (applies identity $\hat{\mathbb{I}}$ to single atoms and $\bar{G}$ to superatoms). \\
$\tilde{\mathcal{G}} = \begin{cases}
\hat{G} \\
\bar{G}
\end{cases}$ & Global operator acting on both single atoms and superatoms. \\
$\vec{\mathcal{G}} = \tilde{\mathcal{T}}\, \bar{\bar{\mathcal{G}}} (\tilde{\mathcal{G}})$ & Wire-gate operator: time- or space-translated gate operation, composed of a translation operator and a conditional gate. \\
\hline\hline
\end{tabular}
\end{table*}

\begin{figure*}[t]
	\centering
	\includegraphics[width=0.95\textwidth]{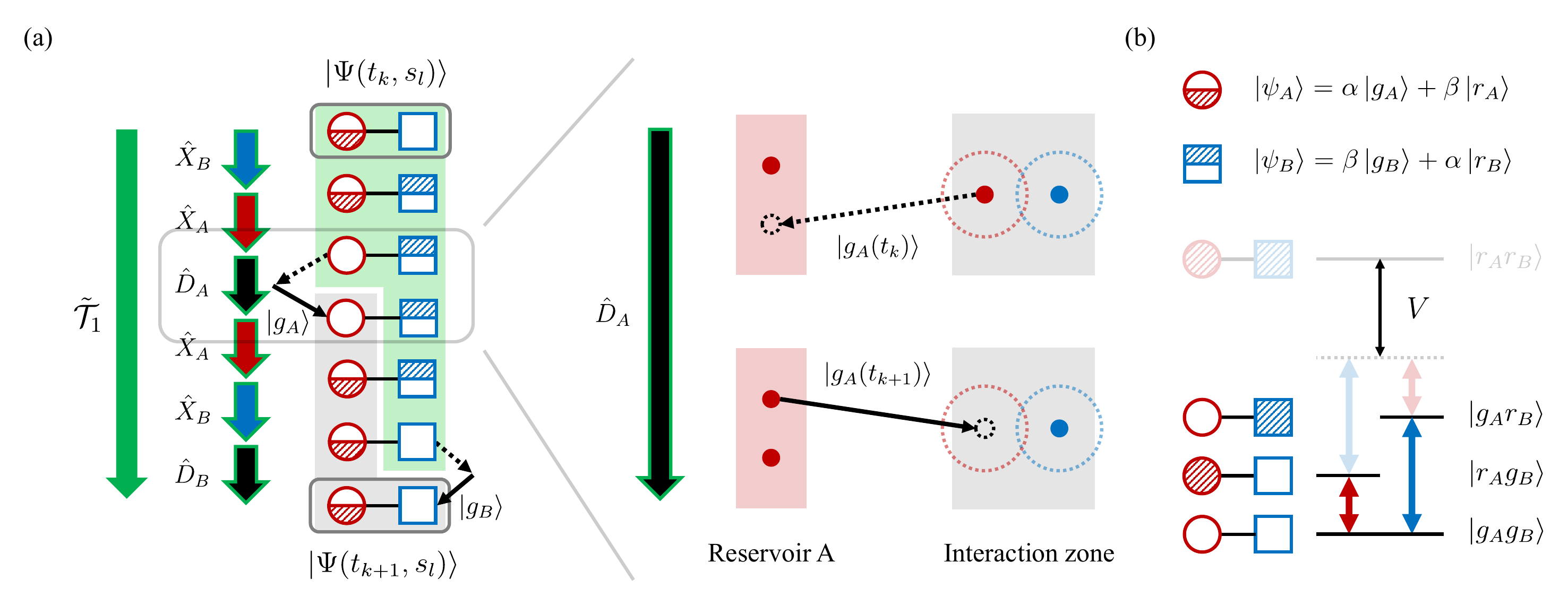}
	\caption{
	\textbf{Temporal mode translation operator $\tilde{\mathcal{T}}_{1}$: single-atom to single-atom transfer.}
	\textbf{(a)} Two successive pulses, $\hat{X}_{A}\hat{X}_{B}$, transfer quantum information from the data qubit (species $A$) to the auxiliary qubit (species $B$) under the Rydberg blockade condition. The replacement operation $\hat{D}_{A}$ removes the $A$ atom at temporal mode $t_{k}$ and loads a fresh $A$ atom at $t_{k+1}$. (Inset: Details of the atom replacement operation. The old data qubit is shuttled to the reservoir, while a new data qubit is delivered to the interaction zone to form the Q-Pair for the next temporal mode.) Iterating this sequence translates the temporal mode from $t_{k}$ to $t_{k+1}$.
	\textbf{(b)} The Rydberg blockade suppresses the doubly excited state $\left| r_{A} r_{B} \right\rangle$, enabling the $\hat{X}_{A}\hat{X}_{B}$ pulse pair to facilitate coherent information flow between the two atoms.
	}
	\label{Fig2}
\end{figure*}

\subsection{Temporal mode translation operation}
At the core of the MAQCY protocol is the quantum translation operator $\tilde{\mathcal{T}}$, which shifts the quantum state $\left| \Psi(t_k, s_l) \right\rangle$ of a Q‑Pair from temporal mode $t_k$ to $t_{k+1}$ (Eq.~\eqref{Eq_TempMux}). The idea of temporal mode translation comes from the CP protocol~\cite{Cesa23}, however, we add the concept of coherent atom shuttling supported by moving tweezer~\cite{Bluvstein22, Bluvstein23, Shaw2024}. Our method can achieve the same functionality as the CP protocol, which requires a quadratic number of atoms proportional to the number of Q-Pairs to guarantee connectivity, while needing to prepare only the Q-Pairs corresponding to a single temporal mode in the time sequence. 

Since data qubits are encoded in either single atoms or superatoms, four types of temporal mode translation operators are defined:
\begin{equation}
\tilde{\mathcal{T}} =
\begin{cases}
\tilde{\mathcal{T}}_1 & \text{for single-atom $\rightarrow$ single-atom,} \\
\tilde{\mathcal{T}}_2 & \text{for single-atom $\rightarrow$ superatom,} \\
\tilde{\mathcal{T}}_3 & \text{for superatom $\rightarrow$ single-atom,} \\
\tilde{\mathcal{T}}_4 & \text{for superatom $\rightarrow$ superatom.}
\end{cases} \label{EqT}
\end{equation}

Each $\tilde{\mathcal{T}}_\nu$ consists of a global bit-flip operation, represented by the Pauli operator $\hat{X}_\mu = \left|g_\mu\right\rangle\left\langle r_\mu\right| + \left|r_\mu\right\rangle\left\langle g_\mu\right|$ with $\mu \in \{A, B\}$, and displacement operators $\hat{D}$ for single atoms and $\bar{D}$ for superatoms,
\begin{subequations}\label{EqD}
\begin{eqnarray}
\hat{D} &=&
\begin{cases}
\hat{D}_A^{\rm in} \, (\hat{D}_A^{\rm out}) & \text{for single-atom $A$ in (out),} \\
\hat{D}_B^{\rm in} \, (\hat{D}_B^{\rm out}) & \text{for single-atom $B$ in (out),}
\end{cases} \label{DHat} \\
\bar{D} &=&
\begin{cases}
\bar{D}_A^{\rm in} \, (\bar{D}_A^{\rm out}) & \text{for superatom $A$ in (out),} \\
\bar{D}_B^{\rm in} \, (\bar{D}_B^{\rm out}) & \text{for superatom $B$ in (out).}
\end{cases} \label{DBar}
\end{eqnarray}
\end{subequations}

Figure~\ref{Fig2}(a) depicts the realization of the temporal mode translation operator $\tilde{\mathcal{T}}_{1}$. The initial state of a Q-Pair in a space-time hybrid mode $(t_k, s_l)$ is
\begin{eqnarray}
\left| \Psi(t_{k},s_{l}) \right\rangle &=& \underbrace{(\alpha\left|g_{A}\right\rangle+\beta\left|r_{A}\right\rangle)}_{\left|\psi_A\right\rangle} \left| g_B \right\rangle,
\end{eqnarray}
with $|\alpha|^{2}+|\beta|^{2}=1$. The superposed state $\left|\psi_A\right\rangle$ is shown as a half-dashed circle, and the ground state $\left|g_B\right\rangle$ as a blue empty rectangle.

The operator $\tilde{\mathcal{T}}_1$ consists of two concatenated quantum gates: $\hat{D}_A\hat{X}_A\hat{X}_B$ followed by $\hat{D}_B\hat{X}_B\hat{X}_A$,
\begin{equation}
\tilde{\mathcal{T}}_1 = \underbrace{\hat{D}_B\hat{X}_B \hat{X}_A} \, \underbrace{\hat{D}_A\hat{X}_A \hat{X}_B}. \label{EqT1}
\end{equation}
Here, $\hat{D}_\mu = \hat{D}_\mu^{\rm in} \hat{D}_\mu^{\rm out}$ denotes an atom replacement operator (Eq.~\eqref{EqD}). Owing to chromatic distinction, each globally applied $\hat{X}_\mu$ pulse selectively addresses the quantum state of atom $\mu$.

The two pulses $\hat{X}_A\hat{X}_B$ in the first half of $\tilde{\mathcal{T}}_1$ evolve the Q-Pair state as
\begin{eqnarray}
\hat{X}_A \hat{X}_B \left|\Psi(t_k,s_l)\right\rangle &=& \left|g_A\right\rangle \underbrace{\left(\beta \left|g_B\right\rangle + \alpha \left|r_B\right\rangle\right)}_{\left|\psi_B\right\rangle}. \label{Eq6}
\end{eqnarray}
The Rydberg blockade permits only transitions $\left|g_Ag_B\right\rangle \rightarrow \left|g_Ar_B\right\rangle$ and $\left|r_Ag_B\right\rangle \rightarrow \left|g_Ag_B\right\rangle$ (Fig.~\ref{Fig2}(b)). Therefore, quantum information flows from the data qubit to the auxiliary qubit.

Then, using the moving tweezer technique, we \textit{discard} the species $A$ atom and \textit{refill} with a newly initialized ground-state atom $A$ from the reservoir (or equivalently via optical pumping to $|g_A\rangle$). This process, which resets the state of atom $A$ and advances its temporal mode, can be described by the following operator:
\begin{eqnarray}
\mathcal{M}_A : |g_A(t_k)\rangle |\psi_B(t_k)\rangle \longmapsto |g_A(t_{k+1})\rangle |\psi_B(t_k)\rangle. \nonumber \\ \label{Eq:MA}
\end{eqnarray}

The inset of Fig.~\ref{Fig2}(a) illustrates the transport of the data qubit via a moving tweezer.
The operation $\mathcal{M}_A$ in Eq.~\eqref{Eq:MA} does not affect the quantum information stored in species $B$. It can be described by a unitary displacement operator $\hat{D}_A = \hat{D}_A^{\rm in} \hat{D}_A^{\rm out}$, satisfying:
\begin{eqnarray}
\hat{D}_A&&\hat{X}_A \hat{X}_B\left|\Psi(t_k,s_l)\right\rangle 
= \nonumber \\
&&\left|g_A(t_{k+1})\right\rangle \left(\beta \left|g_B(t_k)\right\rangle + \alpha \left|r_B(t_k)\right\rangle\right). 
\end{eqnarray}

Similarly, after applying the second half of $\tilde{\mathcal{T}}_1$, the Q-Pair state at temporal mode $t_{k+1}$ becomes
\begin{eqnarray} 
\left|\Psi(t_{k+1},s_l)\right\rangle &=& \tilde{\mathcal{T}}_1 \left|\Psi(t_k,s_l)\right\rangle \nonumber \\
&=& \hat{D}_B\hat{X}_B \hat{X}_A \hat{D}_A\hat{X}_A \hat{X}_B \left|\Psi(t_k,s_l)\right\rangle \nonumber \\
&=& \left|\psi_A(t_{k+1})\right\rangle \left|g_B(t_{k+1})\right\rangle. \label{Telep}
\end{eqnarray}

\begin{figure*}[t]
	\includegraphics[width=0.9\textwidth]{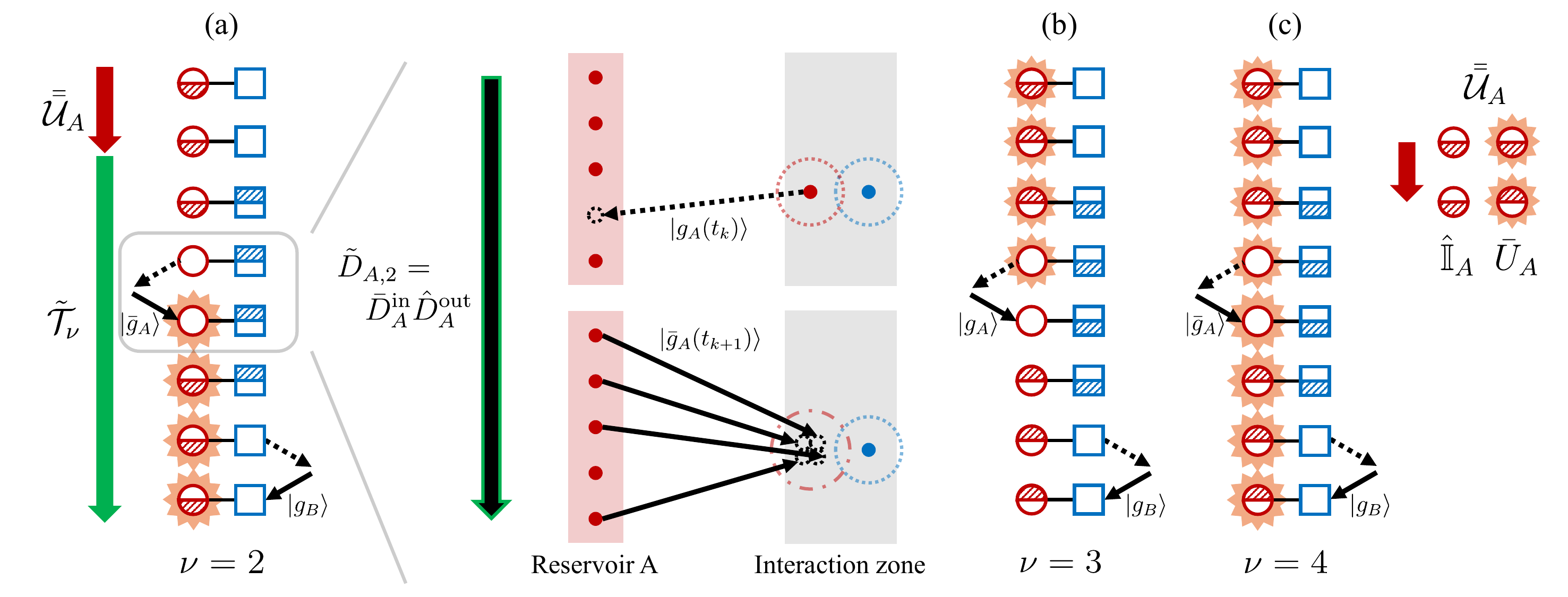}
	\caption{
	\textbf{Superatom-based single-qubit wire-gates} $\vec{\mathcal{U}}_\nu = \tilde{\mathcal{T}}_\nu \bar{\bar{\mathcal{U}}}_A$, $\nu = 2, 3, 4$, with $\bar{\bar{\mathcal{U}}}_A = \bar{X}_A$. 
	\textbf{(a)} For $\nu = 2$, $\bar{\bar{\mathcal{U}}}_A$ is applied to a single-atom $A$ in the Q-Pair at temporal mode $t_k$, and the atom at $t_{k+1}$ that replaces it is a ground-state superatom $A$, denoted by $\left|\bar{g}_A\right\rangle$. 
	\textbf{(b)} For $\nu = 3$, $\bar{\bar{\mathcal{U}}}_A$ is applied to a superatom $A$ at $t_k$, and the atom at $t_{k+1}$ that replaces it is a ground-state single-atom $A$, denoted by $\left|g_A\right\rangle$. 
	\textbf{(c)} For $\nu = 4$, $\bar{\bar{\mathcal{U}}}_A$ is applied to a superatom $A$ at $t_k$, and the atom at $t_{k+1}$ that replaces it is a ground-state superatom $A$, denoted by $\left|\bar{g}_A\right\rangle$. 
	The temporal mode translation operators $\tilde{\mathcal{T}}_\nu$ in (a)-(c) are defined in Eq.~(\ref{EqTG}).
	}
	\label{Fig3}
\end{figure*}

For the other temporal translation operators $\tilde{\mathcal{T}}_2$, $\tilde{\mathcal{T}}_3$, and $\tilde{\mathcal{T}}_4$, we introduce the superatom bit-flip operator $\bar{X}_A = \left|\bar{g}_A\right\rangle\left\langle \bar{r}_A\right| + \left|\bar{r}_A\right\rangle\left\langle \bar{g}_A\right|$, used whenever a superatom is involved. By employing composite global pulses~\cite{Cesa23, Levine19, Levitt1979, Levitt1986, Gulde2003, Schmidt-Kaler2003, Fromonteil2023}, we realize bit-flips simultaneously on both single atoms and superatoms, such that 
\begin{eqnarray}
    \tilde{\mathcal{X}}_A  = \begin{cases}
    \hat{X}_A & \text{for single-atom,}\\
    \bar{X}_A & \text{for superatom.}
\end{cases}
\end{eqnarray}
The composite pulse example for $\tilde{\mathcal{X}}_A$ is given in the Appendix~\ref{App:XX-gate}.

Consequently, all four temporal mode translation operators are generalized as:
\begin{equation}
\tilde{\mathcal{T}}_{\nu} = \hat{D}_B \hat{X}_B \tilde{\mathcal{X}}_A \tilde{D}_{A,\nu} \tilde{\mathcal{X}}_A \hat{X}_B, \quad \nu \in \{1, 2, 3, 4\}, \label{EqTG}
\end{equation}
where:
\begin{equation}
\tilde{D}_{A,\nu} =
\begin{cases}
\hat{D}_A & \text{(single-atom $\rightarrow$ single-atom),} \\
\bar{D}_A^{\rm in} \hat{D}_A^{\rm out} & \text{(single-atom $\rightarrow$ superatom),} \\
\hat{D}_A^{\rm in} \bar{D}_A^{\rm out} & \text{(superatom $\rightarrow$ single-atom),} \\
\bar{D}_A^{\rm in} \bar{D}_A^{\rm out} & \text{(superatom $\rightarrow$ superatom).}
\end{cases}
\label{EqDcases} 
\end{equation}
Note that in Eq.~(\ref{EqTG}), the operators $\hat{X}_B$ and $\hat{D}_B$ are the same as those in Eq.~(\ref{EqT1}). We emphasize that the time-translation operator $\tilde{\mathcal{T}}_\nu$ can be applied globally to each Q-Pair at any spatial mode $s_l$ within the same temporal mode $t_k$.

We note that decoherence during the translation operation $\tilde{\mathcal{T}}$ can constitute a fundamental limitation of the MAQCY protocol. In particular, the displacement operations $\hat{D}$ and $\bar{D}$ consume a significant portion of the total time budget. To mitigate decoherence of the Rydberg state $\left|r\right\rangle$, its population can be temporarily transferred to a more stable state $\left|g'\right\rangle$ during displacement. Further details are discussed in the discussion section (Sec.~\ref{Discussion}).

\subsection{Single-qubit gate}
In the MAQCY protocol, any single-qubit unitary gate is applied just before each of the translation operators $\tilde{\mathcal{T}}_\nu$ in Eq.~(\ref{EqTG}). Note that we use a similar approach to the CP protocol’s single-qubit gate operation~\cite{Cesa23}. However, unlike the CP protocol, in our scheme the single-qubit gate acts only on the data qubit of species $A$ within the Q-Pair.
We construct a global single-qubit gate $\bar{\bar{\mathcal{U}}}_A$ that acts on the Q‑Pair at the hybrid mode $(t_k, s_l)$, selectively affecting superatoms while leaving single atoms unchanged:
\begin{equation} 
\bar{\bar{\mathcal{U}}}_A = 
\begin{cases}  
\hat{\mathbb{I}}_A & \text{for single atoms,} \\
\bar{U}_A & \text{for superatoms.}
\end{cases} 
\label{EqUA}
\end{equation}

By combining the temporal mode translation operators $\tilde{\mathcal{T}}_\nu$ with the single-qubit unitary gate $\bar{\bar{\mathcal{U}}}_A$, we define the unitary wire-gate~\cite{Jeong22}:
\begin{equation}
\vec{\mathcal{U}}_\nu = \tilde{\mathcal{T}}_\nu \bar{\bar{\mathcal{U}}}_A, \quad \nu \in \{1, 2, 3, 4\}. 
\label{EqGW}
\end{equation}
 
Thus, the quantum state $\left|\Psi(t_k,s_l)\right\rangle$ at temporal mode $t_k$ is coherently transformed into the state $\left|\Psi(t_{k+1},s_l)\right\rangle$ at $t_{k+1}$ by applying one of the four unitary wire-gates $\vec{\mathcal{U}}_\nu$, as defined in Eq.~(\ref{EqGW}):
\begin{eqnarray}
\label{EqGU}
&&\left|\Psi(t_{k+1},s_l)\right\rangle = \vec{\mathcal{U}}_\nu \left|\Psi(t_k,s_l)\right\rangle \nonumber \\
&&=
\begin{cases}
\tilde{\mathcal{T}}_1 \hat{\mathbb{I}}_A \left|\psi_A(t_k)\right\rangle \left|g_B(t_k)\right\rangle & \text{for $\nu = 1$}, \\
\tilde{\mathcal{T}}_2 \hat{\mathbb{I}}_A \left|\psi_A(t_k)\right\rangle \left|g_B(t_k)\right\rangle & \text{for $\nu = 2$}, \\
\tilde{\mathcal{T}}_3 \bar{U}_A \left|\bar{\psi}_A(t_k)\right\rangle \left|g_B(t_k)\right\rangle & \text{for $\nu = 3$}, \\
\tilde{\mathcal{T}}_4 \bar{U}_A \left|\bar{\psi}_A(t_k)\right\rangle \left|g_B(t_k)\right\rangle & \text{for $\nu = 4$}.
\end{cases}
\end{eqnarray}
Here, $\left|\psi_A(t_k)\right\rangle$ and $\left|\bar{\psi}_A(t_k)\right\rangle$ denote the quantum states of the single-atom and superatom, respectively, at the end of temporal mode $t_k$.

For $\nu = 1$, the gate $\bar{\bar{\mathcal{U}}}_A = \hat{\mathbb{I}}_A$, i.e., no active gate is applied, and the quantum information encoded in $\left|\Psi(t_k,s_l)\right\rangle$ is transferred coherently to $\left|\Psi(t_{k+1},s_l)\right\rangle$ by $\tilde{\mathcal{T}}_1$ alone, as previously shown in Fig.~\ref{Fig2}. Detailed composite pulse sequences used to implement global unitary gates $\bar{\bar{\mathcal{U}}}_A$$-$such as the bit-flip and Hadamard gates$-$are introduced in the Appendix~\ref{App:CompositePulse}.

Figure~\ref{Fig3} illustrates the remaining three single-qubit wire-gates $\vec{\mathcal{U}}_\nu$ for $\nu = 2$ (a), $\nu = 3$ (b), and $\nu = 4$ (c), using $\bar{U}_A = \bar{X}_A$ as an example.

In case (a) with $\nu = 2$, a single-atom data qubit at $t_k$ is first acted upon by $\bar{\bar{\mathcal{U}}}_A$ and then replaced with a ground-state superatom $\left|\bar{g}_A(t_{k+1})\right\rangle$ (see inset of Fig.~\ref{Fig3}(a)). This process coherently transfers the quantum information to a new superatom state $\left|\bar{\psi}_A(t_{k+1})\right\rangle$.

In contrast, for cases (b) and (c) with $\nu = 3$ and $\nu = 4$, the Q‑Pair initially contains a superatom at $t_k$, which is then mapped to either a single-atom state $\left|\psi_A(t_{k+1})\right\rangle$ or a new superatom state $\left|\bar{\psi}_A(t_{k+1})\right\rangle$, respectively.

\begin{figure}[t]
	\includegraphics[width=0.45\textwidth]{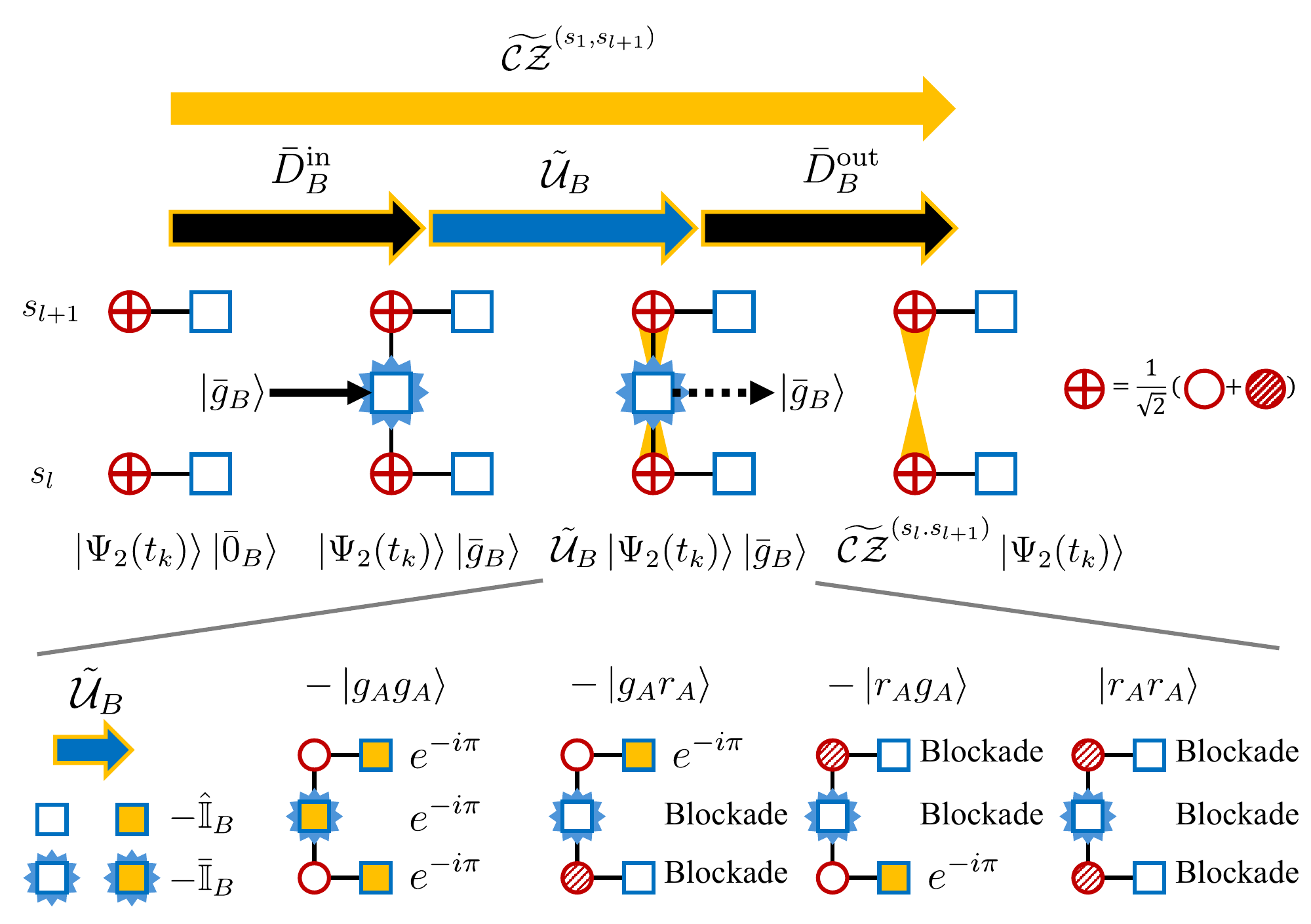}
	\caption{
	\textbf{Two-qubit controlled-Z (CZ) gate implementation.} 
	The displacement operator $\bar{D}_B^{\rm in}$ inserts an auxiliary superatom $B$ between the two Q-Pairs.
	A global pulse sequence $\tilde{\mathcal{U}}_B$ mediates interactions between superatom $B$ and the data qubits of the Q-Pairs.
	Finally, the operator $\bar{D}_B^{\rm out}$ removes superatom $B$, leaving only the two Q-Pairs.
	(Inset) The global operation $\tilde{\mathcal{U}}_B$ assigns a relative phase of $\pi$ to both single-atom and superatom states of species $B$.
	Due to the Rydberg blockade, excitation is forbidden if both data qubits are in the Rydberg state $\left|r_A r_A\right\rangle$.
	This blockade induces a relative $\pi$ phase shift, thereby implementing the CZ gate.
	}
	\label{Fig4}
\end{figure}

\subsection{Two-qubit gate}
For universal and fault-tolerant quantum computing$-$as well as for generating all-to-all connected entanglement$-$controlled gates such as controlled-Z (CZ) and controlled-NOT (CNOT, or CX) gates are essential. In the MAQCY protocol, controlled gates like the CZ gate act on two Q-Pairs, $\left|\Psi(t_k,s_l)\right\rangle$ and $\left|\Psi(t_k,s_{l+1})\right\rangle$, within the same temporal mode $t_k$. To maintain consistency with the wire-gate formalism used for single-qubit gates, the result of the CZ operation must be propagated to the next temporal mode $t_{k+1}$. 

Accordingly, the CZ wire-gate takes the form:
\begin{eqnarray}
\vec{\mathcal{CZ}}_{\mu,\nu}^{(s_l,s_{l+1})} &=& \tilde{\mathcal{T}}^{(s_l)}_\mu \tilde{\mathcal{T}}^{(s_{l+1})}_\nu \widetilde{\mathcal{CZ}}^{(s_{l},s_{l+1})}, \\
&& \mu,\nu \in \{1, 2, 3, 4\}. \nonumber \label{EqGWCZ}
\end{eqnarray}
Here, the two independent translation operators $\tilde{\mathcal{T}}_\mu^{(s_l)}$ and $\tilde{\mathcal{T}}_\nu^{(s_{l+1})}$ act separately on each Q-Pair.

The interaction between the two data qubits is mediated by a superatom of species $B$ (hereafter, superatom $B$). Figure~\ref{Fig4} illustrates the implementation of $\widetilde{\mathcal{CZ}}^{(s_l, s_{l+1})}$ in Eq.~\eqref{EqGWCZ}. This operation consists of in- and out-displacement operators, $\bar{D}_B^{\rm in}$ and $\bar{D}_B^{\rm out}$, which act on superatom $B$, and a global pulse sequence $\tilde{\mathcal{U}}_B$ that drives both Q-Pairs and superatom $B$:
\begin{equation}
\widetilde{\mathcal{CZ}}^{(s_l,s_{l+1})} = \bar{D}_B^{\rm out} \tilde{\mathcal{U}}_B \bar{D}_B^{\rm in}. \label{UACZ}
\end{equation}
In the CP protocol~\cite{Cesa23}, the two-qubit interaction uses a superatom placed between two dual-species atomic wires. By contrast, in MAQCY the entanglement-mediating superatom is mobile, similar to the Q-Pairs themselves.

To illustrate this, consider a representative example. Suppose the initial state of each Q-Pair is given by $|\Psi(t_k,s_l)\rangle = |\Psi(t_k,s_{l+1})\rangle = |\oplus_A\rangle |g_B\rangle$, where $|\oplus_A\rangle = (|g_A\rangle + |r_A\rangle)/\sqrt{2}$ denotes an equal superposition state of the data atom (depicted as a red $\oplus$ in Fig.~\ref{Fig4}). The two-Q-Pair product state is then written as $|\Psi_2(t_k)\rangle = |\Psi(t_k,s_l)\rangle |\Psi(t_k,s_{l+1})\rangle$.

In the first step of $\widetilde{\mathcal{CZ}}^{(s_l,s_{l+1})}$, the superatom $B$ is inserted between the two Q-Pairs. Denoting the absence of superatom $B$ by the vacuum state $\left|\bar{0}_B\right\rangle$, the joint state is initially
\[
\left|\Psi_2(t_k)\right\rangle \left|\bar{0}_B\right\rangle = \left|\oplus_A \oplus_A\right\rangle \left|g_B g_B\right\rangle \left|\bar{0}_B\right\rangle.
\]
Superatom $B$ must interact with the data qubits of the Q-Pairs, but not with their auxiliary qubits. Applying the insertion operation $\bar{D}_B^{\rm in}$ yields:
\begin{equation}
\bar{D}_B^{\rm in} \left|\Psi_2(t_k)\right\rangle \left|\bar{0}_B\right\rangle = \left|\Psi_2(t_k)\right\rangle \left|\bar{g}_B\right\rangle.
\end{equation}

Next, we apply the global composite pulse $\tilde{\mathcal{U}}_B$~\cite{Fromonteil2023}, which depends only on the data-qubit states. This implements a selective phase gate:
\begin{align}
\tilde{\mathcal{U}}_B &=
\begin{cases}
\hat{U}(\frac{\pi}{4},\frac{\pi}{2}) \hat{U}(\pi,0) \hat{U}(\frac{\pi}{2},\frac{\pi}{2}) \hat{U}(\pi,0) \hat{U}(\frac{\pi}{4},\frac{\pi}{2}), \\
\bar{U}(\frac{\pi}{2},\frac{\pi}{2}) \bar{U}(2\pi,0) \bar{U}(\pi,\frac{\pi}{2}) \bar{U}(2\pi,0) \bar{U}(\frac{\pi}{2},\frac{\pi}{2}),
\end{cases}
\nonumber \\
&=
\begin{cases}
-\hat{\mathbb{I}}_B & \text{for single-atom } B, \\
-\bar{\mathbb{I}}_B & \text{for superatom } B.
\end{cases}
\label{EqCZB}
\end{align}
Here, $\hat{U}(\theta,\phi)$ and $\bar{U}(\bar{\theta},\phi)$ are Bloch-sphere rotations defined in Eqs.~\eqref{USin} and \eqref{USup} of the Appendix~\ref{App:CompositePulse}.
The quantities $\theta = \Omega_B \tau$ and $\bar{\theta} = \bar{\Omega}_B \tau = 2\theta$ denote the pulse areas for single-atom and superatom $B$, respectively, and $\phi$ is the phase angle.

This pulse sequence induces a $2\pi$ rotation about the $\hat{y}$-axis for single-atom $B$ and a $6\pi$ rotation for superatom $B$. Due to Rydberg blockade between heterogeneous atoms (Eq.~\eqref{E2b}), all computational basis states except $\left|r_A r_A\right\rangle$ acquire a relative $\pi$ phase via $-\hat{\mathbb{I}}_B$ or $-\bar{\mathbb{I}}_B$, while $\left|r_A r_A\right\rangle$ remains unchanged. 
Thus, the global pulses acts as:
\begin{eqnarray}
\tilde{\mathcal{U}}_B \left|g_B g_B\right\rangle \left|\bar{g}_B\right\rangle \left\langle g_B g_B\right| \left\langle \bar{g}_B\right|: 
\begin{cases}
\left|g_A g_A\right\rangle \mapsto \left|g_A g_A\right\rangle, \\
\left|g_A r_A\right\rangle \mapsto \left|g_A r_A\right\rangle, \\
\left|r_A g_A\right\rangle \mapsto \left|r_A g_A\right\rangle, \\
\left|r_A r_A\right\rangle \mapsto -\left|r_A r_A\right\rangle.
\end{cases}
\end{eqnarray}
In the final step, we apply the displacement operator $\bar{D}_B^{\rm out}$ to remove superatom $B$. 

In summary, the CZ operation $\widetilde{\mathcal{CZ}}^{(s_l,s_{l+1})}$ transforms
\begin{equation}
    |\Psi_2(t_k)\rangle = \left|\oplus_A \oplus_A\right\rangle \left|g_B g_B\right\rangle 
\end{equation} 
into a maximally entangled state:
\begin{eqnarray}
\label{EqUCZ}
&&\widetilde{\mathcal{CZ}}^{(s_l,s_{l+1})} |\Psi_2(t_k)\rangle = 
- \frac{1}{2} \Big[ |g_A g_A\rangle +|g_A r_A\rangle + \nonumber \\ 
&& \qquad \qquad \qquad \qquad |r_A g_A\rangle - |r_A r_A\rangle \Big] \left|g_B g_B\right\rangle.
\end{eqnarray}
This demonstrates that auxiliary superatoms enable entanglement between data qubits while maintaining compatibility with global control.

\begin{figure}[t]
	\includegraphics[width=0.45\textwidth]{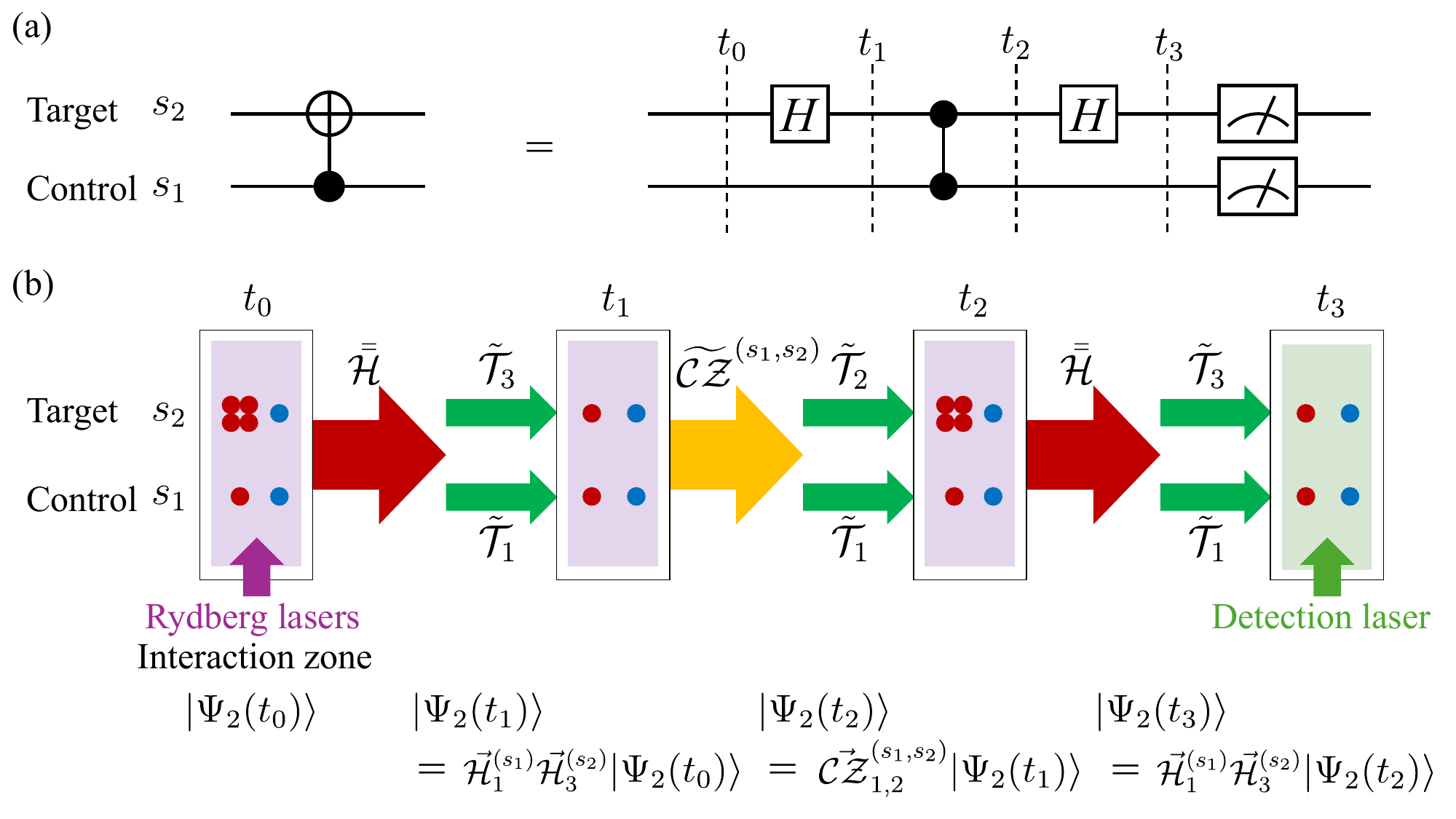}
	\caption{
	\textbf{Two-qubit global CNOT wire-gate $\vec{\mathcal{CX}}^{(s_1,s_2)}_{1,3}$.}
	(a) The CNOT gate is implemented by sandwiching a CZ gate between two Hadamard gates on the target qubit. This is realized using the global Hadamard gate $\bar{\bar{\mathcal{H}}}$ defined in Eq.~(\ref{EqHWG}).
	(b) Experimental realization of the global CNOT wire-gate $\vec{\mathcal{CX}}^{(s_1,s_2)}_{1,3}$ }in the MAQCY protocol.
	\label{Fig5}
\end{figure}

Thus, the global CZ wire-gate $\vec{\mathcal{CZ}}_{\mu,\nu}^{(s_l,s_{l+1})}$ coherently transforms the state $\left|\Psi_2(t_k)\right\rangle$ of two adjacent Q-Pairs in Fig.~\ref{Fig5}(b) into the state at $t_{k+1}$:
\begin{eqnarray}
\left|\Psi_2(t_{k+1})\right\rangle &=& \vec{\mathcal{CZ}}_{\mu,\nu}^{(s_l,s_{l+1})} \left|\Psi_2(t_k)\right\rangle.
\end{eqnarray}

Finally, the CNOT gate can be implemented by combining Hadamard and CZ gates, as shown in Fig.~\ref{Fig5}(a). Figure~\ref{Fig5}(b) shows the corresponding implementation in the MAQCY protocol. The Hadamard wire-gate is defined as
\begin{equation}
\vec{\mathcal{H}}_\nu = \tilde{\mathcal{T}}_\nu \bar{\bar{\mathcal{H}}}, \quad \nu \in \{1,2,3,4\}, \label{EqHWG}
\end{equation}
and is used to construct the global CNOT wire-gate using the global CZ wire-gate as
\begin{eqnarray}
&& \vec{\mathcal{CX}}^{(s_1,s_2)}_{1,3} = \nonumber \\
&&\quad\quad \underbrace{[\vec{\mathcal{H}}^{(s_1)}_1 \vec{\mathcal{H}}^{(s_2)}_3]}_{t_2 \rightarrow t_3}
\underbrace{[\vec{\mathcal{CZ}}^{(s_1,s_2)}_{1,2}]}_{t_1 \rightarrow t_2}
\underbrace{[\vec{\mathcal{H}}^{(s_1)}_1 \vec{\mathcal{H}}^{(s_2)}_3]}_{t_0 \rightarrow t_1}. \label{EqCNOT}
\end{eqnarray}
Here, the single-atom data qubit in hybrid mode $(t_0,s_1)$ acts as the control, and the superatom data qubit in $(t_0,s_2)$ acts as the target, as shown in Fig.~\ref{Fig5}(b). Note that each bracketed block, such as $[\vec{\mathcal{H}}^{(s_1)}_1 \vec{\mathcal{H}}^{(s_2)}_3]$, corresponds to a global pulse. As an example, we present the case $(\mu,\nu) = (1,3)$, but this choice is not restrictive.

\begin{figure*}[t]
	\includegraphics[width=0.9\textwidth]{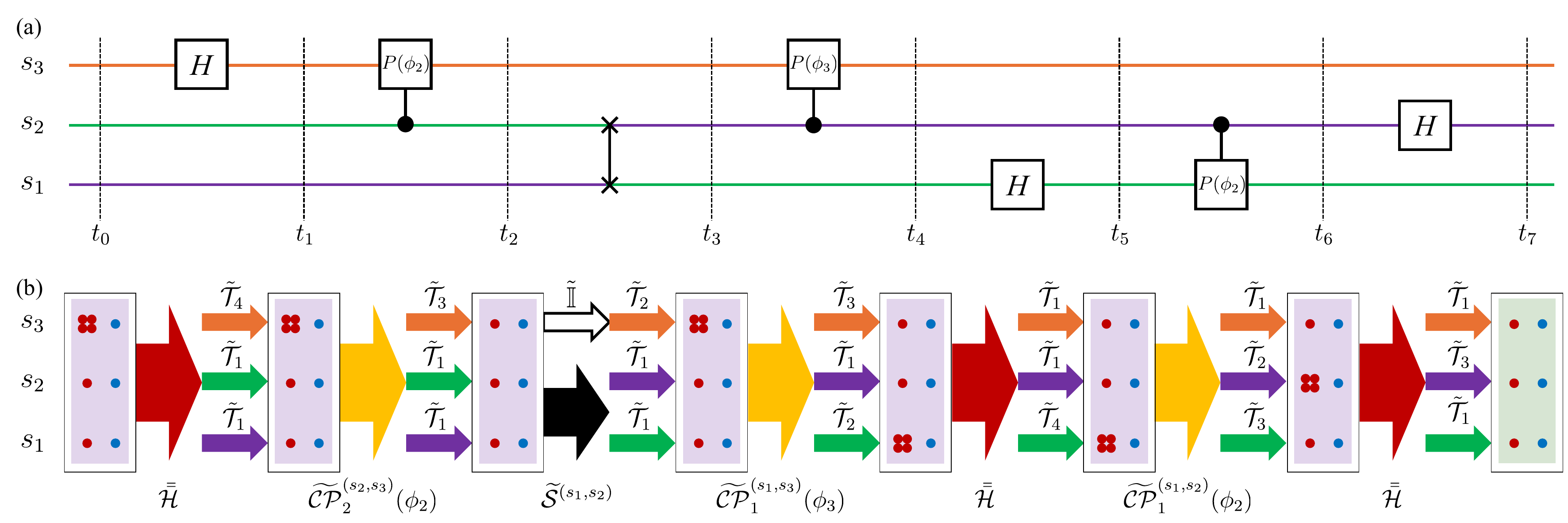}
	\caption{
	\textbf{Three-qubit quantum Fourier transform circuit.}
	\textbf{(a)} Quantum circuit model of a three-qubit quantum Fourier transform (QFT).
	\textbf{(b)} Experimental realization of the three-qubit QFT using the MAQCY protocol.
	Red arrows: single-qubit Hadamard operations; orange, green, and purple  }arrows: temporal mode translation operations; yellow arrow: two-qubit C-Phase operation; black arrow: atom SWAP; outlined arrow: identity operation.
	\label{Fig6}
\end{figure*}

\subsection{Universal quantum computation}
As demonstrated above, our MAQCY protocol enables the implementation of arbitrary single-qubit wire-gates $\vec{\mathcal{U}}_\nu$, as well as two-qubit entanglement generation via the CZ wire-gate $\vec{\mathcal{CZ}}^{(s_l,s_{l+1})}_{\mu,\nu}$ and the CNOT wire-gate $\vec{\mathcal{CX}}^{(s_l,s_{l+1})}_{\mu,\nu}$. By combining these building blocks, all Clifford quantum gates can be implemented. 

As an example, Fig.~\ref{Fig6} illustrates the implementation of a three-qubit quantum Fourier transform (QFT) using our MAQCY architecture. The QFT circuit additionally requires a two-qubit SWAP wire-gate $\vec{\mathcal{S}}^{(s_l,s_m)}$, defined in Eq.~(\ref{SwapWG}), and a global controlled-phase (C-Phase) wire-gate $\vec{\mathcal{CP}}_l^{(s_l,s_{l+1})}(\phi_q)$, defined in Eq.~\eqref{EqCP} in the Appendix~\ref{App:Cgate}, with discrete phase $\phi_q = 2\pi/2^q$, $q \in \mathbb{Z}$.
The global C-Phase wire-gate $\vec{\mathcal{CP}}^{(s_l,s_{l+1})}_{l}(\phi_q)$ can be decomposed into two single-qubit phase gates and two CNOT gates. Detailed constructions are provided in the Appendix~\ref{App:Cgate}.

The SWAP gate $\tilde{\mathcal{S}}^{(s_l,s_m)}$ is essential for realizing all-to-all connectivity in the MAQCY protocol. It enables two distant Q-Pairs to interact by swapping their spatial positions $s_l$ and $s_m$ through atom movement:
\begin{equation}
    \tilde{\mathcal{S}}^{(s_l,s_m)} = \begin{cases}
        |\Psi(s_l)\rangle \mapsto |\Psi(s_m)\rangle,\\
        |\Psi(s_m)\rangle \mapsto |\Psi(s_l)\rangle.
    \end{cases} \label{Swap}
\end{equation}

When combined with temporal mode translation operators, the SWAP gate becomes a wire-gate compatible with the MAQCY protocol:
\begin{eqnarray}\label{SwapWG}
\vec{\mathcal{S}}_{\mu,\nu}^{(s_l,s_{m})} &=& \tilde{\mathcal{T}}^{(s_l)}_\mu \tilde{\mathcal{T}}^{(s_{m})}_\nu\widetilde{\mathcal{S}}^{(s_{l},s_{m})}, \\
&&\mu,\nu \in \{1, 2, 3, 4\}. \nonumber
\end{eqnarray}

In Fig.~\ref{Fig6}, we initialize the three-qubit QFT by loading a superatom $A$ and a single-atom $B$ into the space-time position $(t_0, s_3)$, and single-atom pairs $(A,B)$ into $(t_0, s_2)$ and $(t_0, s_1)$, thereby forming three Q-Pairs.

The first operation is a global Hadamard wire-gate $\bar{\bar{\mathcal{H}}}$ (large red arrow) at $t_0$, which creates an equal superposition in the Q-Pair at $s_3$ only. Then, three temporal mode translation operators are applied: 
$\tilde{\mathcal{T}}_4$ (orange), $\tilde{\mathcal{T}}_1$ (green), and $\tilde{\mathcal{T}}_1$ (purple), respectively, advancing all three Q-Pairs to temporal mode $t_1$.

Next, the QFT circuit proceeds with a well-defined sequence of global single- and two-qubit wire-gates. At temporal mode \( t_1 \), a controlled-phase gate \(\vec{\mathcal{CP}}_2^{(s_2,s_3)}(\phi_2)\) is applied between the second and third Q-Pairs. This is followed by a SWAP gate \(\vec{\mathcal{S}}^{(s_1,s_2)}\), executed between \( t_2 \) and \( t_3 \), to exchange the quantum information encoded in the first and second Q-Pairs. Subsequently, at \( t_3 \), another controlled-phase gate \(\vec{\mathcal{CP}}_2^{(s_2,s_3)}(\phi_3)\) is applied between the same pair. At \( t_4 \), a global Hadamard gate \(\bar{\bar{\mathcal{H}}}\) is performed on the Q-Pair located at position \( (t_4, s_1) \), followed by a third controlled-phase gate \(\vec{\mathcal{CP}}_2^{(s_2,s_3)}(\phi_2)\) at \( t_5 \). Finally, at \( t_6 \), a Hadamard operation \(\bar{\bar{\mathcal{H}}}\) is applied to the Q-Pair at \( (t_6, s_2) \). 

At the conclusion of these operations, all three Q-Pairs are measured at \( t_7 \) to complete the implementation of the quantum Fourier transform within the MAQCY protocol.

\section{Experimental platform}
\begin{figure}[t]
	\includegraphics[width=0.45\textwidth]{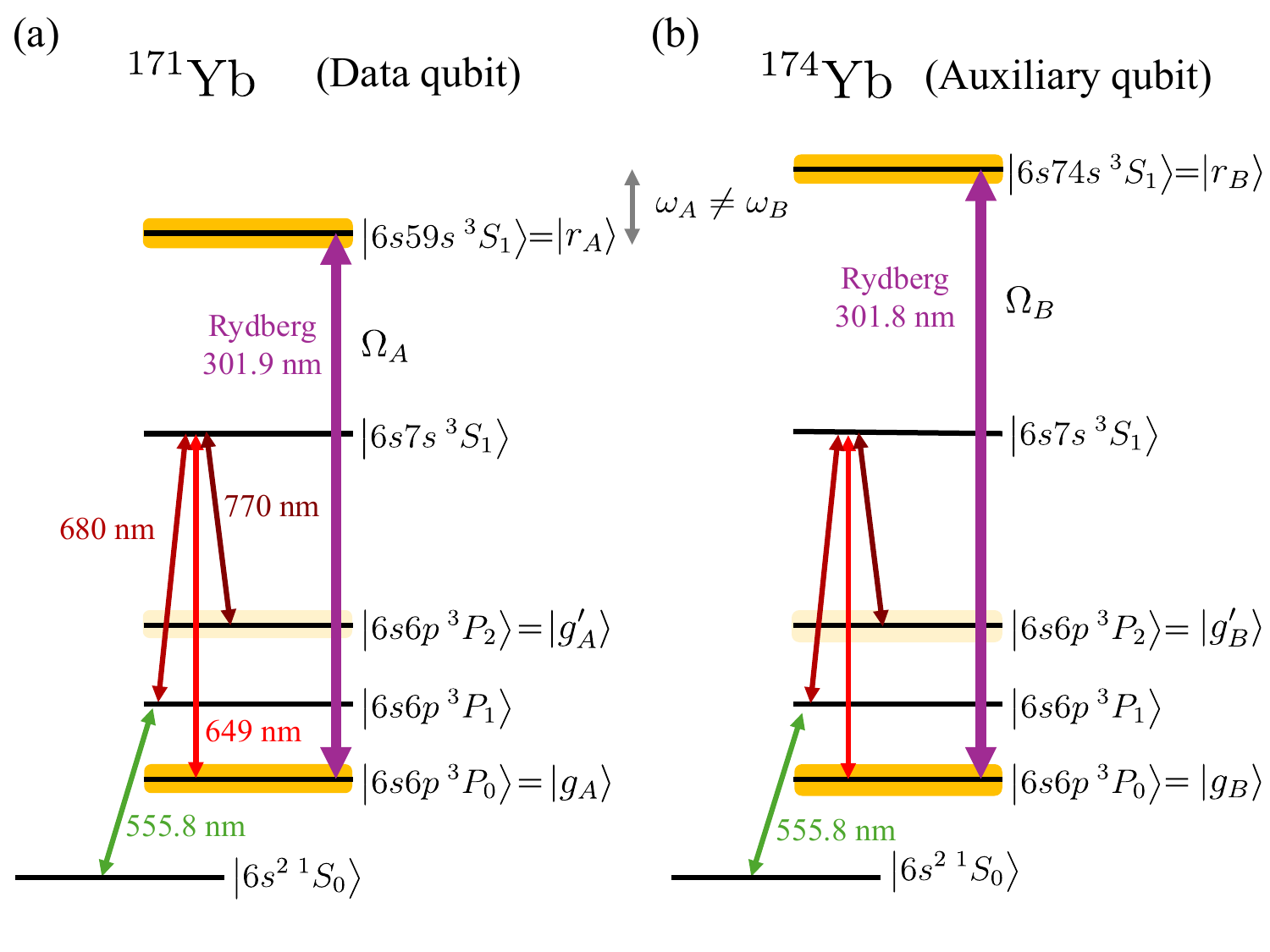}
	\caption{Relevant energy-level structures and electron configurations of the $^{171}$Yb (a) and $^{174}$Yb (b) isotopes. In our MAQCY protocol, a single $^{171}$Yb atom (species $A$) with resonant frequency $\omega_A$ and Rabi frequency $\Omega_A$ serves as the data qubit, while a single $^{174}$Yb atom (species $B$) with resonant frequency $\omega_B$ ($\ne \omega_A$) and Rabi frequency $\Omega_B$ serves as the auxiliary qubit. Population measurement is achieved by detecting green scattered photons following optical pumping to the $\left|6s7s~{}^3S_1\right>$ state.
    } \label{Fig7}
\end{figure}

As a concrete example of the experimental realization of the MAQCY protocol, we consider an atomic platform based on dual ytterbium isotopes, \(^{171}\mathrm{Yb}\) (fermionic) and \(^{174}\mathrm{Yb}\) (bosonic), confined in optical tweezers, as recently demonstrated~\cite{Nakamura24}. These isotopes offer the most detailed spectroscopic data for Rydberg states of atomic ytterbium to date~\cite{Peper2025}. In our Q‑Pair scheme, the \(^{171}\mathrm{Yb}\) atoms serve as data qubits (species \(A\)), while the \(^{174}\mathrm{Yb}\) atoms act as auxiliary qubits (species \(B\)). Figure~\ref{Fig7} illustrates the relevant energy-level structures and electron configurations.

The ground states of the data and auxiliary qubits are defined as $\left|g_A\right>=\left|6s6p~{}^3P_0, F=1/2, m_F=1/2\right>$ and $\left|g_B\right>=\left|6s6p~{}^3P_0\right>$, respectively-optical clock states with long lifetimes ($\sim$100~s) and high resilience to ambient magnetic field fluctuations~\cite{McGrew18}. The Rydberg states are defined as $\left|r_A\right>=\left|6s\,n s~{}^3S_1, F=1/2, m_F=1/2\right>$ for data qubits and $\left|r_B\right>=\left|6s\,n's~{}^3S_1\right>$ for auxiliary qubits. Transitions $|{g_\mu}\rangle\leftrightarrow|{r_\mu}\rangle$ are driven at around 302~nm with Rabi frequencies $\Omega_\mu$. By choosing appropriate principal quantum numbers (e.g., $n=59$, $n'=74$), one achieves a frequency separation $\lvert\omega_A - \omega_B\rvert/(2\pi) \gtrsim 329$~GHz, sufficient to enable species-selective global driving~\cite{Peper2025, Wilson2022}.

Both Yb isotopes can be trapped in the same 759.2~nm optical tweezers$-$a magic wavelength for the ${}^1S_0 - {}^3P_0$ clock transition~\cite{Brown2017,Hohn2023}. Atoms are initialized in $\left|g_{\mu}\right>$ using a three-photon transition via $\left|6s6p~{}^3P_1\right>$ and $\left|6s7s~{}^3S_1\right>$ intermediate states, using lasers at 555.8~nm, 680~nm, and 649~nm. This avoids requiring large magnetic fields for direct excitation to ${}^3P_0$ in ${}^{174}\mathrm{Yb}$. Before the displacement operations, Rydberg-state populations $\left|r_{\mu}\right>$ are de-excited to metastable states $\left|g'_\mu\right> = \left|6s6p~{}^3P_2\right>$ via 326~nm light, circumventing anti-trapping effects. These $\left|g'_\mu\right>$ states are subsequently shuttled using auxiliary tweezers at 532~nm, which stably confine the ${}^3P_2$ state.

Since quantum information resides solely in the data qubits, only the final states of the data atoms must be measured. Nonetheless, species-selective readout is possible due to the multi-GHz difference in ${}^1S_0 - {}^3P_1$ transition frequencies between ${}^{171}$Yb and ${}^{174}$Yb~\cite{Nakamura24} isotopes. 

Ground-state population measurement ($|g_A\rangle$, $|g_B\rangle$) is performed by resonantly pumping atoms to the absolute ground state $\left|6s^2~{}^1S_0\right>$ while detecting the green fluorescence from the 555.8~nm transition with a single-photon-sensitive detector such as an electron multiplying charge-coupled device
(EMCCD).

We note that residual population in the auxiliary Rydberg state acts as a signature for auto-ionization fidelity and can thus be interpreted as an erasure channel~\cite{Wu2022, Scholl2023, Ma2023}. Furthermore, spatially separated readout-zone~\cite{Bluvstein23} enable selective qubit readout, which is independent with Q-Pairs in the interaction zone.

\section{DISCUSSION} \label{Discussion}
The MAQCY protocol achieves scalability by leveraging temporal modes as an additional computational degree of freedom. This architecture requires only $\mathcal{O}(N)$ physical atoms to enable all-to-all qubit connectivity for $N$ data qubits. In contrast to architectures that rely on local optical addressing$-$susceptible to beam-pointing instabilities and thermal motion$-$our scheme employs global control pulses only, enhancing robustness and reducing hardware complexity.

The bit-flip probability is quantified as $P_d = N' \Gamma \tau$~\cite{Lechner2015}, where $N'$ denotes the number of required atoms, $\Gamma$ is the decoherence rate, and $\tau$ is the total operation time. For the CP protocol~\cite{Cesa23}, where $N' = \mathcal{O}(N^2)$ and $\tau = \mathcal{O}(NP)$ (with $P$ denoting circuit depth), the bit-flip probability scales as $P_d = \mathcal{O}(N^3 P)$. In contrast, MAQCY reduces this to $P_d = \mathcal{O}(N^2)$ by virtue of $N' = \mathcal{O}(N)$ and $\tau = \mathcal{O}(N)$, independent of $P$.

Using the optical clock and Rydberg states of ytterbium, a single-qubit gate $\hat{X}_\mu$ with Rabi frequency $\Omega = 2\pi \times 10~\mathrm{MHz}$ yields a gate time of $t_g = \pi/\Omega = 0.05~\mu$s. With Rydberg lifetime $\Gamma^{-1} \sim 60~\mu$s~\cite{Ma2023, Peper2025}, the associated error per gate is $p=\Gamma t_g / 2 = 0.0004$~\cite{Wu2022}. Mid-circuit erasure detection can mitigate this error by filtering out unwanted decay events~\cite{Madjarov2020}. Experiments have reported 33~\% error suppression using such techniques~\cite{Ma2023}, corresponding to a single-qubit gate fidelity of $F_X = 0.9997$.

Note that, under spontaneous decay, the MAQCY protocol provides a superlinear fidelity advantage compared to the single-atom approach, as demonstrated in the Appendix~\ref{App:Noise}. While the single-atom protocol exhibits an average fidelity of $\langle F \rangle = 1 - \mathcal{O}(p)$ for a single time-translation operation, the MAQCY protocol can achieve $\langle F^{\rm QP} \rangle = 1 - \mathcal{O}(p^2)$ for the same time-translation operation for a single Q-Pair. The calculation details are provided in the Appendix~\ref{App:Noise}.

Further fidelity enhancement is achieved by offloading quantum information from fragile Rydberg states to long-lived metastable states $|{g'_\mu}\rangle = |{6s6p~{}^3P_2}\rangle$ during atomic motion. With $\Gamma_{g'}^{-1} \sim 15$~s~\cite{Mishra2001}, the memory fidelity for $\sim500~\mu$s shuttling is $F_{D,\Gamma} = 0.99997$. Assuming the same gate fidelity $F_{X'} = F_X$ for transitions between $|{g'_\mu}\rangle$ and $|{r_\mu}\rangle$, and a displacement fidelity $F_{D,\mathrm{mov}} \gtrsim 0.995$~\cite{Bluvstein22, Shaw2024}, the overall fidelity of the temporal translation operator $\tilde{\mathcal{T}}_1$ (involving four $\hat{X}_\mu$, four $\hat{X}'_\mu$, and two displacement operations) becomes
\begin{equation}
F_\mathcal{T} = F_X^8 (F_{D,\Gamma} F_{D,\mathrm{mov}})^2 \approx 0.99.
\end{equation}
If the movement fidelity improves to $F_{D,\mathrm{mov}} = 0.999$, $F_\mathcal{T}$ can reach 0.995, surpassing the surface-code threshold~\cite{Fowler2012}.

Note here that we assume the single-qubit fidelity of $F_X = 0.9997$; however, the state-of-the-art ytterbium platform reported $F_X = 0.9990$~\cite{Ma2023}. To reach the error-correction threshold, $F_X = 0.9996$ is required under the assumption of $F_{D,\mathrm{mov}} = 0.999$, which has been demonstrated on an alkali atom-based neutral-atom platform~\cite{Xia2015}.

Moreover, mid-circuit measurement can be used to check if atoms are well-prepared before gate operations begin. If a vacancy is detected, we can use atom shuttling with moving tweezers to refill it.
An atom's motional state can be initialized by sideband cooling. The recently developed erasure cooling method~\cite{Shaw2025}, which couples an atom's internal and motional states, provides a way to perform mid-circuit measurement and correction of a non-ground motional state. 

The composite pulse sequences used for MAQCY gates are not unique and may benefit from further optimization using pulse-shaping techniques~\cite{Jandura2022} or time-optimal control via Hamilton-Jacobi-Bellman methods~\cite{Fromonteil2024}. Although we have focused on a planar layout, MAQCY can be naturally extended to three-dimensional configurations~\cite{kusano2025}. Additionally, ground-metastable qudit encodings~\cite{Jia2024} unique to $^{171}$Yb may further enhance capacity and error resilience.

Finally, MAQCY could be implemented using a single atomic species, e.g., 
AEAs
$-$where both ground and Rydberg states are trapped simultaneously via engineered optical potentials~\cite{Madjarov2020, Wilson2022, Topcu2014}. Floquet engineering~\cite{Zhao2023} offers another promising route for gate construction. Alternatively, a fully single-species $^{171}$Yb system using distinct nuclear spin states and polarization-selective laser coupling provides another viable implementation path.

\section{Conclusions.}
We have proposed and analyzed a scalable quantum computing architecture$-$MAQCY$-$based on globally driven, dual-species neutral-atom arrays with space-time multiplexing. The fundamental computational unit in MAQCY is the Q-Pair: a co-trapped, dual-species qubit pair that enables global gate operations and temporal mode translation.

Our protocol supports arbitrary single-qubit gates, controlled-Z, controlled-phase, and controlled-NOT gates, as well as composite quantum circuits such as the three-qubit quantum Fourier transform. All these operations are realized using only global laser pulses and spatiotemporal reconfiguration of atoms.

Crucially, the physical qubit overhead of MAQCY scales linearly, $\mathcal{O}(N)$, in contrast to the $\mathcal{O}(N^2)$ scaling in previous proposals. By combining fast, programmable atom rearrangement, long-lived metastable memory states, and mid-circuit erasure detection, our protocol achieves high fidelity under experimentally feasible parameters. Once the fidelity of atomic transport improves$-$which is currently the primary bottleneck$-$MAQCY will emerge as a competitive and scalable platform for fault-tolerant universal quantum computing.

\begin{acknowledgments}
We are grateful for stimulating discussions with Francesco Cesa and Joonhee Choi. This work was supported by the National Research Foundation of Korea (NRF-2022M3K4A1094781).
\end{acknowledgments}

\bibliographystyle{quantum}
\bibliography{MAQCY.bib}

\appendix

\section{Composite Pulses for Global Unitary Operations} \label{App:CompositePulse}

In the MAQCY protocol, we utilize only \textit{resonant global driving} to implement arbitrary single- and two-qubit operations. This type of driving permits rotations on the Bloch sphere about an axis in the $xy$-plane, defined by the unit vector $\hat{n} = (\cos\phi, \sin\phi, 0)$. The angle $\phi$ sets the rotation axis on the equator of the Bloch sphere, while the rotation angles $\theta$ and $\bar{\theta}$ determine the magnitude of the rotation for the single-atom and superatom, respectively, as shown below:
\begin{equation}
\begin{cases}
\hat{U}(\theta,\phi), & \text{for a single-atom,} \\
\bar{U}(\bar{\theta},\phi), & \text{for a superatom,}
\end{cases} \label{Eq:UbarDef}
\end{equation}

The resulting unitary transformation of the Bloch vector for single-atom is given by 
\begin{align}
&\hat{U}(\theta,\phi) = e^{-i\frac{\theta}{2}\hat{n}\cdot\hat{\bf{\sigma}}} =\exp\left[ -i\frac{\theta}{2} \left( \cos\phi\,\hat{X} + \sin\phi\,\hat{Y} \right) \right] \nonumber \\
&= \cos\left(\frac{\theta}{2}\right) \hat{\mathbb{I}} - i \sin\left(\frac{\theta}{2}\right) \left( \cos\phi\,\hat{X} + \sin\phi\,\hat{Y} \right), \label{USin}
\end{align}
where $\hat{\sigma}$ is the Pauli spin operator, $\theta = \Omega \tau$ is the pulse area, $\hat{X} = |g\rangle\langle r| + |r\rangle\langle g|$, $\hat{Y} = -i|g\rangle\langle r| + i|r\rangle\langle g|$, and $\hat{\mathbb{I}} = |g\rangle\langle g| + |r\rangle\langle r|$. In practice, one can design a pulse with arbitrary Rabi frequency $\Omega$, duration $\tau$, and phase $\phi$.

On the other hand, superatoms evolve differently under the same resonant pulse due to their collectively enhanced Rabi frequency. Without loss of generality, we consider a superatom of size $M = 4$, for which the collective Rabi frequency is $\bar{\Omega} = 2\Omega$. For the same $\Omega$, $\tau$, and $\phi$, the unitary evolution of the superatom qubit becomes 
\begin{align}
&\bar{U}(\bar{\theta}, \phi) = e^{-i\frac{\bar{\theta}}{2}\hat{n}\cdot\hat{\sigma}} = \exp\left[ -i\frac{\bar{\theta}}{2} \left( \cos\phi\,\bar{X} + \sin\phi\,\bar{Y} \right) \right] \nonumber \\
&= \cos\left(\frac{\bar{\theta}}{2}\right) \bar{\mathbb{I}} - i \sin\left(\frac{\bar{\theta}}{2}\right) \left( \cos\phi\,\bar{X} + \sin\phi\,\bar{Y}\right), \label{USup}
\end{align}
where the pulse area is $\bar{\theta} = 2\theta$, and the Pauli operators for the superatom are defined analogously: $\bar{X} = |\bar{g}\rangle\langle \bar{r}| + |\bar{r}\rangle\langle \bar{g}|$, $\bar{Y} = -i|\bar{g}\rangle\langle \bar{r}| + i|\bar{r}\rangle\langle \bar{g}|$, and $\bar{\mathbb{I}} = |\bar{g}\rangle\langle \bar{g}| + |\bar{r}\rangle\langle \bar{r}|$.

As an example, a resonant pulse of duration $\tau = \pi/\Omega = 2\pi/\bar{\Omega}$ yields different outcomes for single-atom and superatom qubits. The single-atom undergoes a $\pi$-rotation about the X-axis on the Bloch sphere, which implements a full bit-flip gate ($\hat{U}(\pi, 0) = \exp(-i\pi \hat{X}/2)$). By contrast, due to the collective enhancement, the superatom experiences a $2\pi$-rotation on the Bloch sphere ($\bar{U}(2\pi, 0) = -\bar{\mathbb{I}}$), which returns the state to its original position and corresponds to a global phase with no net bit-flip.

By exploiting these \textit{differential Rabi rotations}, we demonstrate three composite global pulse-based operations: $\tilde{\mathcal{X}}$, $\bar{\bar{\mathcal{X}}}$, and $\bar{\bar{\mathcal{H}}}$. These operations are described in detail below.

\begin{figure}[t]
\centerline{\includegraphics[width=0.45\textwidth]{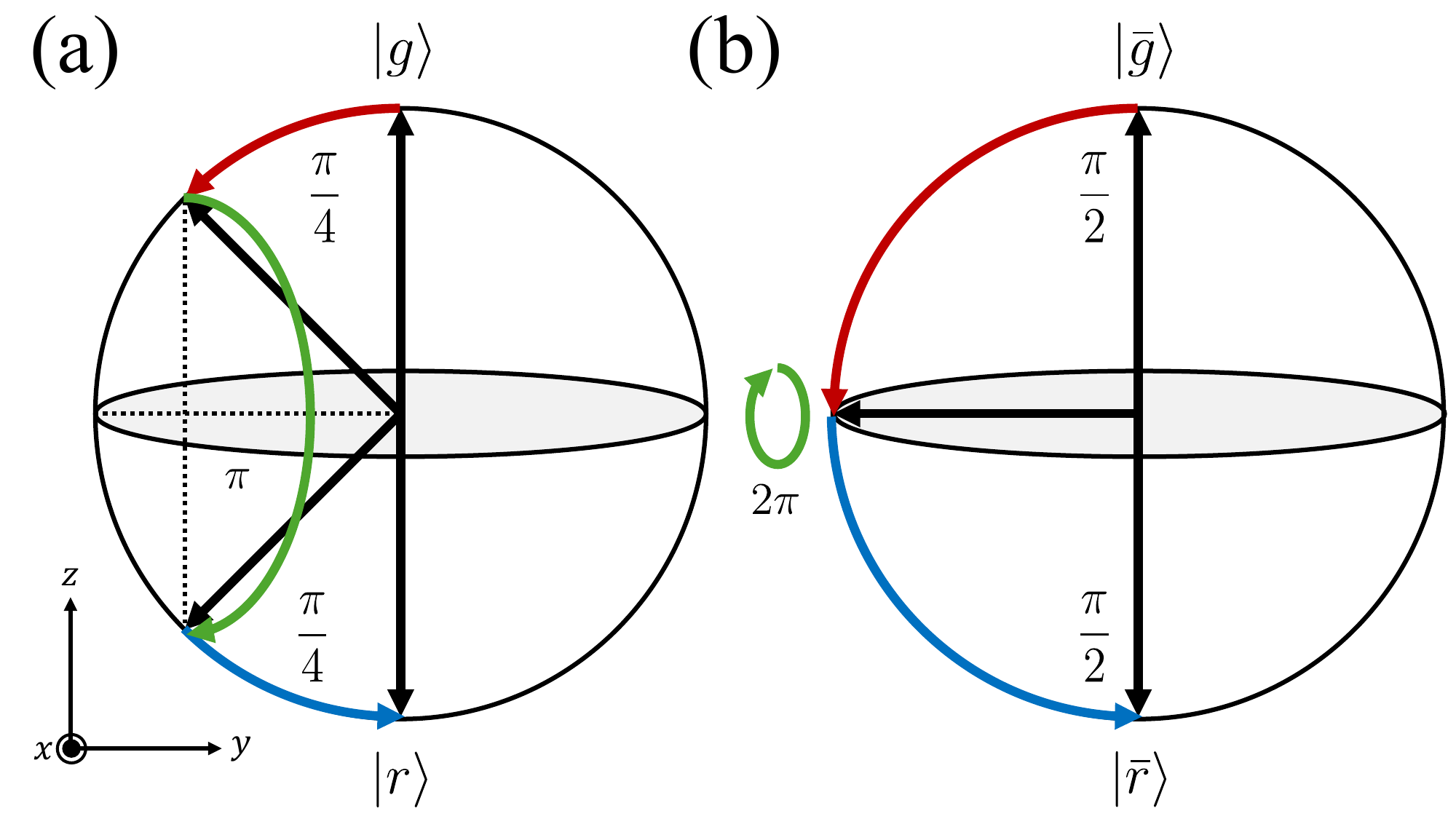}}
\caption{Illustration of composite pulse sequences: (a) for the $\hat{X}_A$ operator acting on a single-atom qubit and (b) for the $\bar{X}_A$ operator acting on a superatom qubit, showing sequential Bloch vector rotations on the surface of the Bloch sphere for initial states $|g\rangle$ and $|\bar{g}\rangle$, respectively.}
	\label{Fig8}
\end{figure}

\subsection{Global bit-flip gates $\tilde{\mathcal{X}}$} \label{App:XX-gate}
Figure~\ref{Fig8} illustrates the sequential Bloch vector rotations generated by the global bit-flip operator $\tilde{\mathcal{X}}$, defined in Eq.~\eqref{EqUX} below. 
This operator is applied both (a) to a single-atom data qubit and (b) to a superatom data qubit.

As discussed in Ref.~\cite{Cesa23}, $\tilde{\mathcal{X}}$ can be implemented using a three-pulse composite sequence:
\begin{align} \label{EqUX}
\tilde{\mathcal{X}} &=\nonumber 
\begin{cases}
\hat{U}\left(\frac{\pi}{4},0\right)\hat{U}\left(\pi,\frac{\pi}{2}\right)\hat{U}\left(\frac{\pi}{4},0\right) = -i \hat{Y},\\
\bar{U}\left(\frac{\pi}{2},0\right)\bar{U}\left(2\pi,\frac{\pi}{2}\right)\bar{U}\left(\frac{\pi}{2},0\right) = i\bar{X},
\end{cases} 
\nonumber \\
&= 
\begin{cases}
      {\buildrel -\hat{P}(\pi) \over \Longrightarrow} \; \hat{X}, & \text{for single-atom}, \\
      {\buildrel -i\bar{P}(0) \over \Longrightarrow} \; \bar{X}, & \text{for superatom},
\end{cases}
\end{align}
where the notation ${\buildrel \cdot \over \Longrightarrow}$ indicates that a subsequent phase gate is applied. Specifically, $\hat{P}_A(\phi) = |g_A\rangle\langle g_A| + e^{i\phi} |r_A\rangle\langle r_A|$ denotes a single-atom phase gate, and $\bar{P}_A(\phi) = |\bar{g}_A\rangle\langle \bar{g}_A| + e^{i\phi} |\bar{r}_A\rangle\langle \bar{r}_A|$ denotes its superatom counterpart. For example, $-\hat{P}(\pi)(-i\hat{Y}) = \hat{X}$ and $\bar{P}(0)(i\bar{X}) = \bar{X}$.

As seen in Fig.~\ref{Fig8} and Eq.~\eqref{EqUX}, the composite pulse sequence $\tilde{\mathcal{X}}$ yields the same bit-flip operator for both types of data qubits: $\hat{X}$ for single atoms and $\bar{X}$ for superatoms, up to a global phase factor. Importantly, a relative phase difference is introduced between the two systems, even when the logical operations are identical. By appropriately controlling this phase, one can implement a conditional superatom phase gate:
\begin{equation} \label{EqPG}
\bar{\bar{\mathcal{P}}}(\phi) = 
\begin{cases} 
\hat{\mathbb{I}}_A, & \text{for single-atom}, \\
\bar{P}(\phi), & \text{for superatom}.
\end{cases}
\end{equation}

To incorporate this into circuit-level operations, we define a phase wire-gate by adapting the wire-gate framework as introduced in Eq.~\eqref{EqGW} to a phase gate operation:
\begin{equation} \label{EqPhaseWire}
\vec{\mathcal{P}}_\nu(\phi) = \tilde{\mathcal{T}}_\nu\, \bar{\bar{\mathcal{P}}}_A(\phi), \quad \nu \in \{1, 2, 3, 4\},
\end{equation}
where $\tilde{\mathcal{T}}_\nu$ denotes a temporal-mode translation operator for Q-Pair index $\nu$.

\begin{figure}[t]
\centerline{\includegraphics[width=0.45\textwidth]{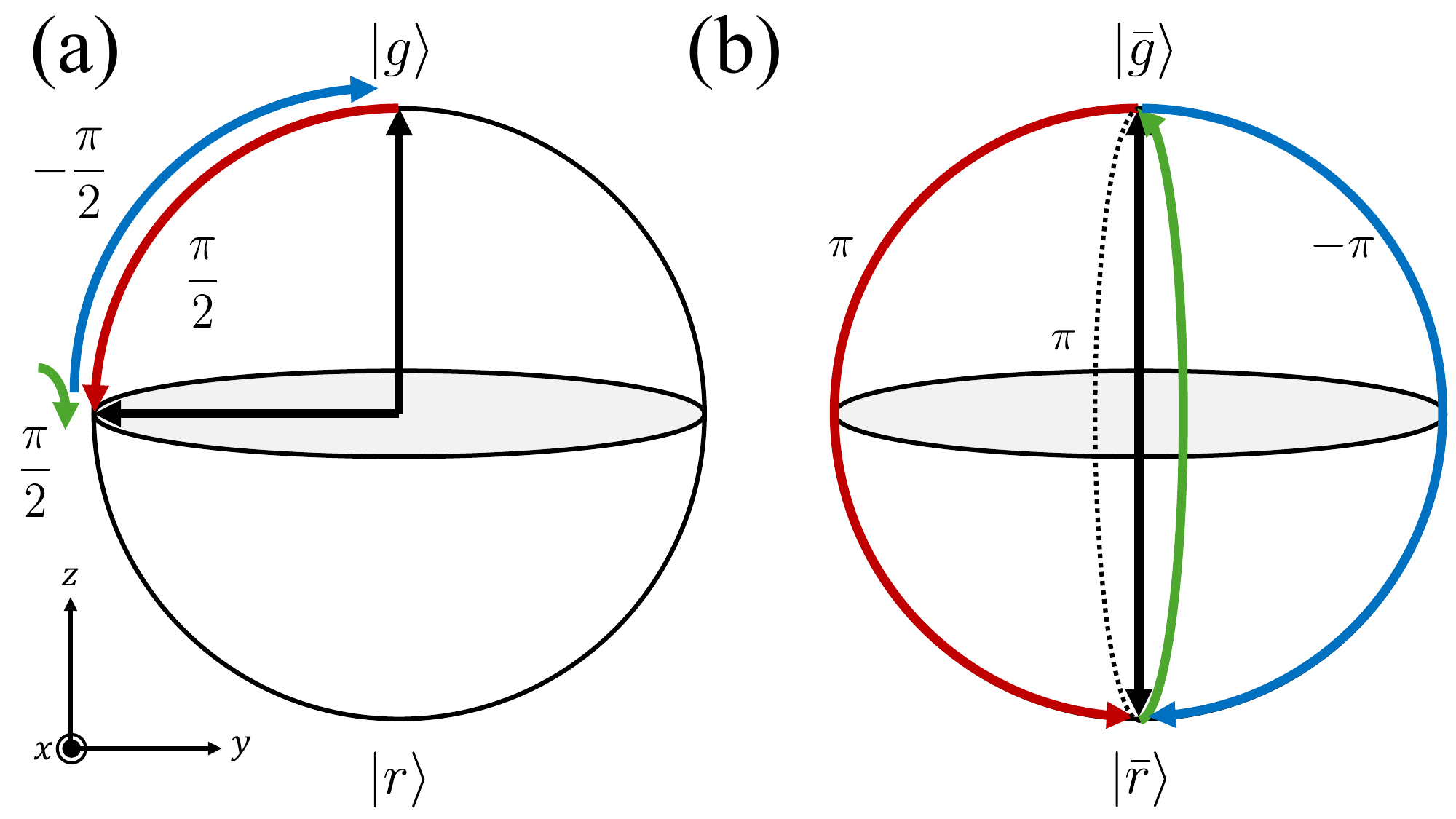}}
	\caption{Illustration of the composite pulse sequences for the global $\bar{\bar{\mathcal{X}}}$ operator applied to a single-atom data qubit $|g\rangle$ (a) and a superatom data qubit $|\bar{g}\rangle$ (b), represented on the Bloch sphere.}
	\label{Fig9}
\end{figure}

\subsection{Global superatom bit-flip gate $\bar{\bar{\mathcal{X}}}$}
To implement the MAQCY protocol, we require a global superatom bit-flip gate $\bar{\bar{\mathcal{X}}}$, which is uniquely designed to flip the superatom data qubit while leaving the single-atom data qubit unaffected. This selectivity enables differential control over qubit species within a global pulse framework. The composite pulse sequence implementing $\bar{\bar{\mathcal{X}}}$ is defined as follows:
\begin{align} \label{EqSX}
\bar{\bar{\mathcal{X}}} &=\nonumber 
\begin{cases}
\hat{U}^\dagger\left(\frac{\pi}{2}, 0\right) 
\hat{U}\left(\frac{\pi}{2}, \frac{\pi}{2}\right)
\hat{U}\left(\frac{\pi}{2}, 0\right) 
= \begin{pmatrix}
e^{i\frac{\pi}{8}} & 0 \\
0 & e^{-i\frac{\pi}{8}}
\end{pmatrix}, \\
\bar{U}^\dagger\left(\pi, 0\right) 
\bar{U}\left(\pi, \frac{\pi}{2}\right)
\bar{U}\left(\pi, 0\right) 
= \begin{pmatrix}
0 & 1 \\
-1 & 0
\end{pmatrix},
\end{cases} 
\nonumber \\
&= 
\begin{cases}
{\buildrel e^{-i\frac{\pi}{8}}\hat{P}(\frac{\pi}{4}) \over \Longrightarrow} \; \hat{\mathbb{I}}, & \text{for single-atom}, \\
{\buildrel \bar{P}(-\pi) \over \Longrightarrow} \; \bar{X}, & \text{for superatom}.
\end{cases}
\end{align}

In this expression, the single-atom pulse sequence results in a global phase rotation, equivalent to the identity operation up to a phase factor. In contrast, the superatom pulse sequence implements a logical $\bar{X}$ gate (bit-flip) exactly. The notation ${\buildrel \cdot \over \Longrightarrow}$ again indicates that an appropriate phase gate follows to complete the desired logical transformation. 

Figure~\ref{Fig9} shows the Bloch vector dynamics under this gate for both (a) a single-atom data qubit initialized in $|g\rangle$, and (b) a superatom data qubit initialized in $|\bar{g}\rangle$. The global gate $\bar{\bar{\mathcal{X}}}$ is thus a key component enabling selective logical control in space-time multiplexed quantum architectures such as MAQCY.

\subsection{Global Hadamard gate $\bar{\bar{\mathcal{H}}}$}
In the MAQCY protocol, a global Hadamard gate $\bar{\bar{\mathcal{H}}}$ is designed to operate identically on both single-atom and superatom data qubits. This gate can be implemented using a sequence of three resonant pulses, similar to the composite construction used for the global bit-flip gate. The composite pulse sequence defining $\bar{\bar{\mathcal{H}}}$ is given by:
\begin{align} \label{EqHD}
\bar{\bar{\mathcal{H}}} &= \nonumber 
\begin{cases}
\hat{U}^\dagger\left(\frac{\pi}{2}, 0\right)
\hat{U}\left(\frac{\pi}{4}, \frac{\pi}{2}\right)
\hat{U}\left(\frac{\pi}{2}, 0\right)
= e^{i\frac{\pi}{8}}, \\[4pt]
\bar{U}^\dagger\left(\pi, 0\right)
\bar{U}\left(\frac{\pi}{2}, \frac{\pi}{2}\right)
\bar{U}\left(\pi, 0\right)
= \frac{1}{\sqrt{2}} \left( \bar{\mathbb{I}} + i\bar{Y} \right),
\end{cases} 
\nonumber \\
&=
\begin{cases}
{\buildrel \hat{P}(-\frac{\pi}{4}) \over \Longrightarrow} \; \hat{\mathbb{I}}, & \text{for single-atom}, \\
{\buildrel \bar{P}(\pi) \over \Longrightarrow} \; \bar{H}, & \text{for superatom}.
\end{cases}
\end{align}

Here, the single-atom sequence yields a global phase factor equivalent to the identity operation, while the superatom sequence implements the desired Hadamard gate \(\bar{H}\) up to a correctable phase. As before, the symbolic arrow notation indicates the application of a final phase gate to achieve the target logical operation. 

\begin{figure}[t]
    \centerline{\includegraphics[width=0.45\textwidth]{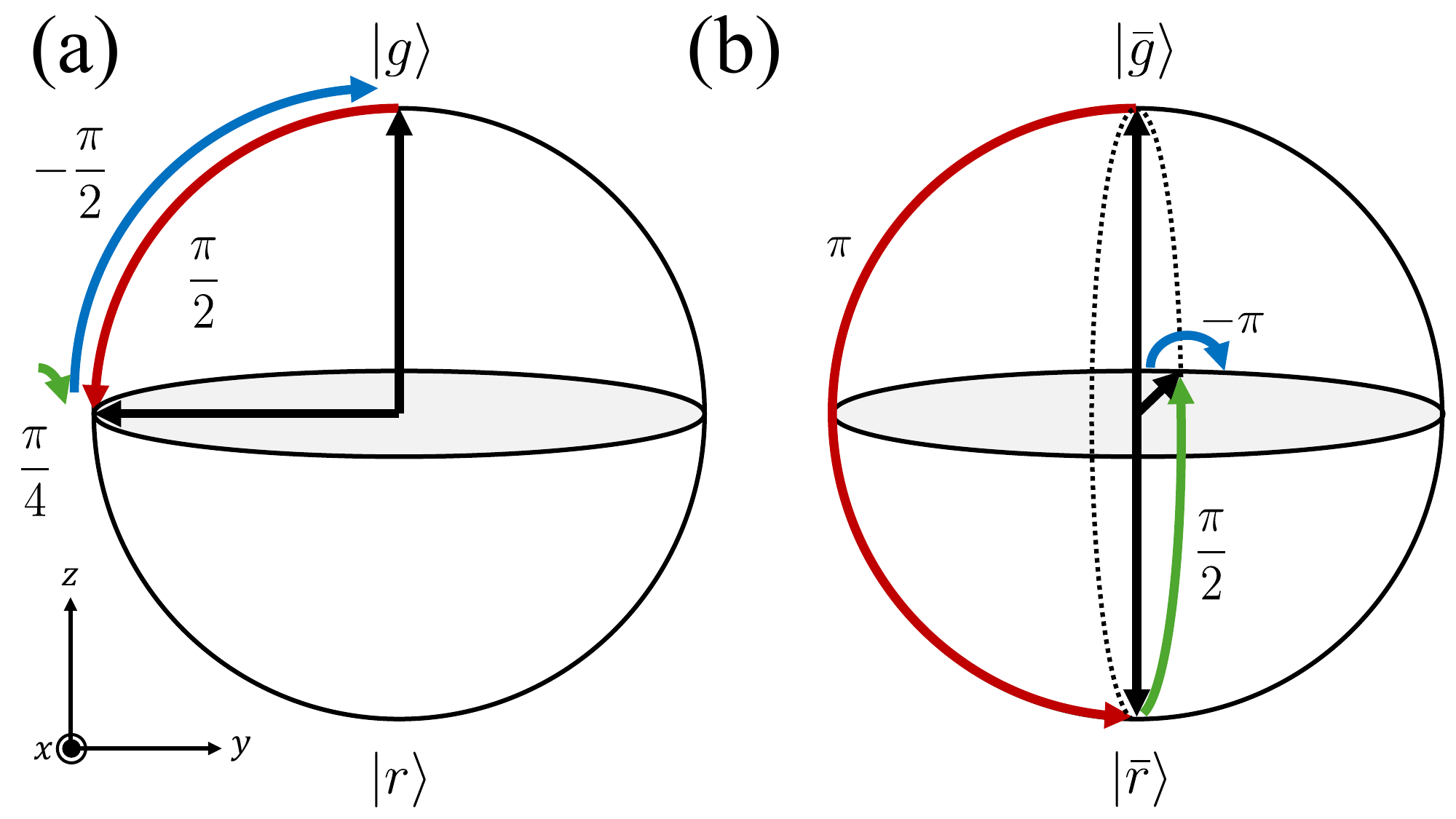}}
	\caption{Illustration of the composite pulse sequences for the global superatom $\bar{\bar{\mathcal{H}}}$ operator acting on a single-atom data qubit $|g\rangle$ (a) and a superatom data qubit $|\bar{g}\rangle$ (b) on the Bloch sphere.}
	\label{Fig10}
\end{figure}

Figure~\ref{Fig10} illustrates the resulting Bloch vector rotations on the sphere induced by \(\bar{\bar{\mathcal{H}}}\) for (a) a single-atom data qubit initially in $|g\rangle$ and (b) a superatom data qubit initially in $|\bar{g}\rangle$. This global Hadamard gate plays a central role in enabling universal logic for hybrid-encoded quantum states in MAQCY.

\section{Controlled-Phase Gate} \label{App:Cgate}

\begin{figure*}[t]
    \centering
    \includegraphics[width=0.9\textwidth]{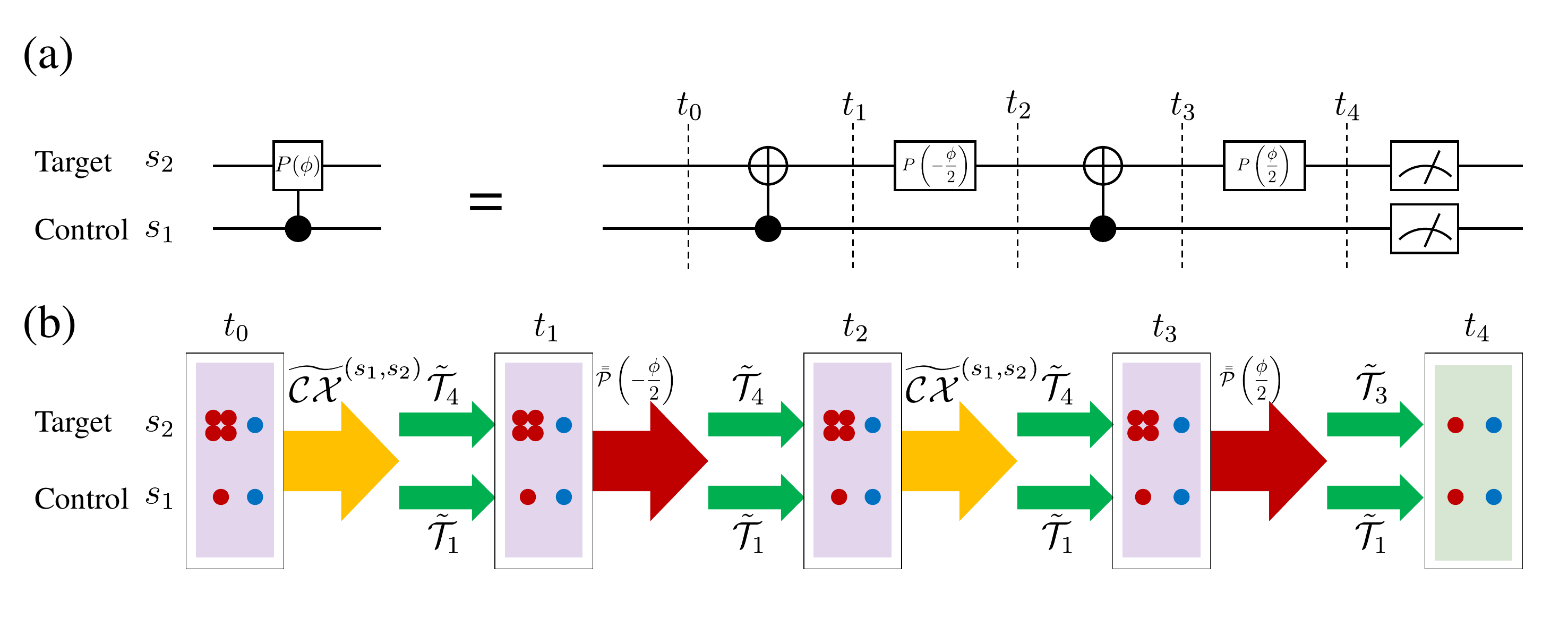} 
    \caption{
    \textbf{Two-qubit global C-Phase wire-gate.}
    \textbf{(a)} The C-Phase gate can be constructed from two CNOT operations interleaved with single-qubit phase gates. 
    \textbf{(b)} Experimental realization of the global two-qubit C-Phase wire-gate \(\vec{\mathcal{CP}}_1^{(s_1,s_2)}(\phi)\) in the MAQCY protocol, showing space-time hybrid control across temporal modes.
    }
    \label{FigCP}
\end{figure*}

The C-Phase gate can be implemented by combining two controlled-NOT (CNOT) gates with two single-qubit phase gates defined in Eq.~\eqref{EqPG}. Figure~\ref{FigCP} illustrates the implementation of the C-Phase gate between two neighboring Q-Pairs located at the hybrid modes \((t_0, s_1)\) and \((t_0, s_2)\), where the atom at site \(s_1\) serves as the control qubit and the atom at site \(s_2\) acts as the target qubit. The complete protocol for the C-Phase gate consists of four sequential steps:

First, a CNOT wire-gate \(\vec{\mathcal{CX}}^{(s_1,s_2)}_{1,4}\) is applied, where the control qubit is $s_1$.
Although the CNOT operation spans four temporal modes, it is abstracted here as a single wire-gate mapping \(t_0 \rightarrow t_1\).

Second, a single-qubit phase gate is applied to the target qubit at \(s_2\), utilizing the wire-gate definition in Eq.~\eqref{EqPhaseWire}. To prepare for the next CNOT operation at \(t_2\), the control qubit at \(s_1\) must be in a single-atom configuration, while the target at \(s_2\) must be in a superatom configuration. Considering the necessary temporal-mode translations, this phase gate is realized over the interval \(t_1 \rightarrow t_2\) by the wire-gate sequence:
\begin{equation}
\left[\vec{\mathcal{P}}^{(s_1)}_1 (-\phi/2), \; \vec{\mathcal{P}}^{(s_2)}_4 (-\phi/2)\right]. 
\end{equation}

Third, the CNOT gate \(\vec{\mathcal{CX}}^{(s_1,s_2)}_{1,4}\) is applied again at \(t_2 \rightarrow t_3\), this time with modified mode translations to accommodate the final phase gate.

Finally, at \(t_4\), both Q-Pairs return to single-atom configurations, and the inverse phase gate is applied as:
\begin{equation}
\left[\vec{\mathcal{P}}^{(s_1)}_1 (\phi/2), \; \vec{\mathcal{P}}^{(s_2)}_1 (\phi/2)\right].
\end{equation}

In summary, the full sequence implementing the global controlled-phase wire-gate is given by:
\begin{align} \label{EqCP}
&\vec{\mathcal{CP}}_1^{(s_1,s_2)} (\phi) =
\underbrace{[\vec{\mathcal{P}}^{(s_1)}_1(\phi/2) \; \vec{\mathcal{P}}^{(s_2)}_1(\phi/2)]}_{t_3 \rightarrow t_4}
 \underbrace{[\vec{\mathcal{CX}}^{(s_1,s_2)}_{1,4}]}_{t_2 \rightarrow t_3}\nonumber \\
&\qquad \times
\underbrace{[\vec{\mathcal{P}}^{(s_1)}_1(-\phi/2) \; \vec{\mathcal{P}}^{(s_2)}_4(-\phi/2)]}_{t_1 \rightarrow t_2}\underbrace{[\vec{\mathcal{CX}}^{(s_1,s_2)}_{1,4}]}_{t_0 \rightarrow t_1}.
\end{align}

This controlled-phase wire-gate \(\vec{\mathcal{CP}}_1^{(s_1,s_2)} (\phi)\) serves as a fundamental building block for multi-qubit quantum algorithms, such as the quantum Fourier transform (QFT) demonstrated in Fig.~\ref{Fig6}.

\section{Noise Modeling using Kraus Formalism} \label{App:Noise}

In this section, we analyze the noise behavior of the temporal mode translation operator $\tilde{\mathcal{T}}_1$, 
using it as a representative example. The single-qubit gate time$-$assumed identical for species $A$ and $B$$-$is set to $t_g = \pi/\Omega = 0.05~\mu$s, which is much shorter than the Rydberg decay time $\Gamma^{-1} \sim 60~\mu$s. This disparity justifies the use of a gate-independent noise model:
\begin{equation}
\exp[(\mathcal{L}_{\text{gate}} + \mathcal{L}_{\text{noise}}) t] \approx \exp[\mathcal{L}_{\text{noise}} t] \exp[\mathcal{L}_{\text{gate}} t],
\end{equation}
where $\mathcal{L}_{\text{gate}}$ and $\mathcal{L}_{\text{noise}}$ denote the Lindbladian superoperators for gate operations and noise processes, respectively. To quantify the noise impact, we employ the Kraus operator-sum representation (OSR) formalism~\cite{Lidar19}.

To apply this to our system, we consider a Q-Pair at time $t_k$, where the data qubit $A$ is in the state $|\psi(t_k)\rangle = \alpha|g_A\rangle + \beta|r_A\rangle$, and the auxiliary qubit $B$ is initialized in $|g_B\rangle$. The corresponding density operator is:
\begin{equation}
    \rho^{\rm QP}(t_k) = \rho_A(t_k) \otimes |g_B\rangle\langle g_B|,
\end{equation}
with $\rho_A(t_k) = |\psi(t_k)\rangle\langle\psi(t_k)|$. The application of $\hat{X}_B$ transforms this state to:
\begin{align} \label{EqDen}
    \rho^{\rm QP}(t_k + t_g^-) &= \hat{X}_B \rho^{\rm QP}(t_k) \hat{X}_B^{\dagger} \nonumber \\
    &= \begin{pmatrix}
        0 & 0 & 0 \\
        0 & |\alpha|^2 & \alpha \beta^* \\
        0 & \alpha^* \beta & |\beta|^2
    \end{pmatrix},
\end{align}
where $t_g^-$ indicates the time immediately before noise is introduced. Under the PXP constraint, the component $\left|r_Ar_B\right\rangle$ is neglected. The dominant noise considered is amplitude damping due to spontaneous decay.

The amplitude-damping channel $\mathcal{E}_{\text{noise}}$ is modeled using the Kraus operators:
\begin{align}
\mathcal{E}_{\text{noise}}[\rho] &= K_0 \rho K_0^\dagger + K_1 \rho K_1^\dagger, \\
K_0 &= \begin{pmatrix} 1 & 0 \\ 0 & \sqrt{1 - p} \end{pmatrix}, \quad
K_1 = \begin{pmatrix} 0 & \sqrt{p} \\ 0 & 0 \end{pmatrix},
\end{align}
where $p = \Gamma t_g / 2$~\cite{Wu2022}. For the bipartite Q-Pair system, the noise channel becomes:
\begin{equation}
\mathcal{E}_{\text{noise}} [\rho^{\rm QP}] = \sum_{i,j \in \{0,1\}} (K^{(A)}_i \otimes K^{(B)}_j)\, \rho^{\rm QP}\, (K^{(A)}_i \otimes K^{(B)}_j)^\dagger.
\end{equation}
The corresponding Kraus operator combinations yield:
\begin{subequations}
\begin{align}
K^{(A)}_0 \otimes K^{(B)}_0 &= \begin{pmatrix} 1 & 0 & 0 \\ 0 & \sqrt{1-p} & 0 \\ 0 & 0 & \sqrt{1-p} \end{pmatrix},  \\
K^{(A)}_0 \otimes K^{(B)}_1 &= \begin{pmatrix} 0 & \sqrt{p} & 0 \\ 0 & 0 & 0 \\ 0 & 0 & 0 \end{pmatrix}, \\
K^{(A)}_1 \otimes K^{(B)}_0 &= \begin{pmatrix} 0 & 0 & \sqrt{p} \\ 0 & 0 & 0 \\ 0 & 0 & 0 \end{pmatrix},  \\
K^{(A)}_1 \otimes K^{(B)}_1 &= 0.
\end{align}
\end{subequations}

Thus, applying this noise channel to Eq.~\eqref{EqDen}, we obtain:
\begin{align}
&\rho^{\rm QP}(t_k + t_g^+) = \mathcal{E}_{\text{noise}} [\rho^{\rm QP}(t_k + t_g^-)] \nonumber \\
&\qquad   = \begin{pmatrix}
p & 0 & 0 \\
0 & |\alpha|^2 (1-p) & \alpha\beta^* (1-p) \\
0 & \alpha^* \beta (1-p) & |\beta|^2 (1-p)
\end{pmatrix}.
\end{align}

Now, after the application of $\hat{X}_A$, the state becomes:
\begin{align}
&\rho^{\rm QP}(t_k + t_1^-) = \hat{X}_A\, \rho^{\rm QP}(t_k + t_g^+) \, \hat{X}_A^\dagger \nonumber \\
&\qquad = \begin{pmatrix}
|\beta|^2(1-p) & \alpha^*\beta (1-p) & 0 \\
\alpha\beta^* (1-p) & |\alpha|^2 (1-p) & 0 \\
0 & 0 & p
\end{pmatrix}.
\end{align}

In turn, after the second amplitude-damping process, the state evolves:
\begin{align}
\rho^{\rm QP}(t_k + t_1^+) &= \mathcal{E}_{\text{noise}} [\rho^{\rm QP}(t_k + t_1^-)]\nonumber \\
&= 
\begin{pmatrix}
\mathcal{A} & \mathcal{B} & 0 \\
\mathcal{B}^* & \mathcal{C} & 0 \\
0 & 0 & \mathcal{D}
\end{pmatrix},
\end{align}
with:
\begin{subequations}
\begin{align}
\mathcal{A} &= p + (1-p)^2 |\beta|^2, \\
\mathcal{B} &= \alpha^* \beta (1-p)^{3/2}, \\
\mathcal{C} &= |\alpha|^2 (1-p)^2, \\
\mathcal{D} &= p(1-p).
\end{align}
\end{subequations}

In the middle, the erasure process~\cite{Wu2022, Scholl2023, Ma2023} removes the $\left|r_A\right\rangle$ component:
\[
\rho^{\rm QP}(t_k + t_1^+) \rightarrow 
\frac{1}{\mathcal{A} + \mathcal{C}}
\begin{pmatrix}
\mathcal{A} & \mathcal{B} & 0 \\
\mathcal{B}^* & \mathcal{C} & 0 \\
0 & 0 & 0
\end{pmatrix}.
\]

This implies the data qubit state $\rho_A(t_k)$ is transferred to the auxiliary qubit as:
\begin{equation}
\rho^{\rm QP}_B(t_k + t_1) =
\begin{pmatrix}
\mathcal{A}' & \mathcal{B}' \\
\mathcal{B}'^* & \mathcal{C}'
\end{pmatrix},
\end{equation}
where:
\begin{subequations}
\begin{align}
\mathcal{A}' &= \frac{p + (1-p)^2 |\beta|^2}{1 - p + p^2}, \\
\mathcal{B}' &= \frac{\alpha^* \beta (1-p)^{3/2}}{1 - p + p^2}, \\
\mathcal{C}' &= \frac{|\alpha|^2 (1-p)^2}{1 - p + p^2}.
\end{align}
\end{subequations}

Finally, after applying the second half of $\tilde{\mathcal{T}}_1$, we recover the data qubit state at $t_{k+1}$:
\begin{equation}
\rho^{\rm QP}_A(t_{k+1}) =
\begin{pmatrix}
\mathcal{A}'' & (\mathcal{B}'')^* \\
\mathcal{B}'' & \mathcal{C}''
\end{pmatrix},
\end{equation}
with:
\begin{subequations}
\begin{align}
\mathcal{A}'' &= \frac{p + (1-p)^2 \mathcal{C}'}{1 - p + p^2}, \\
\mathcal{B}'' &= \frac{\mathcal{B}' (1-p)^{3/2}}{1 - p + p^2}, \\
\mathcal{C}'' &= \frac{\mathcal{A}' (1-p)^2}{1 - p + p^2}.
\end{align}
\end{subequations}

Thus, the final state after $\tilde{\mathcal{T}}_1$ is:
\begin{align}
&\rho^{\rm QP}_A(t_{k+1}) = \nonumber \\
&\begin{pmatrix}
\frac{p(1 - p + p^2) + |\alpha|^2 (1-p)^4}{(1 - p + p^2)^2} &
\frac{\alpha \beta^* (1-p)^3}{(1 - p + p^2)^2} \\
\frac{\alpha^* \beta (1-p)^3}{(1 - p + p^2)^2} &
\frac{(1-p)^2[p + (1-p)^2|\beta|^2]}{(1 - p + p^2)^2}
\end{pmatrix}.
\end{align}

For comparison, the state of a single atom experiencing decay over a total time of $4t_g$ is:
\begin{equation}
\rho_A(t_{k+1}) =
\begin{pmatrix}
|\alpha|^2 + 4p|\beta|^2 & \alpha^* \beta \sqrt{1 - 4p} \\
\alpha \beta^* \sqrt{1 - 4p} & (1 - 4p)|\beta|^2
\end{pmatrix}.
\end{equation}

As the final step, we calculate the fidelities with respect to the initial state \(\left|\psi_A(t_k)\right\rangle = \alpha|g_A\rangle + \beta|r_A\rangle\) as follows:
\begin{align}
F_A &= \langle \psi_A(t_k) | \rho_A(t_{k+1}) | \psi_A(t_k) \rangle \nonumber \\
&= 1 - 4p|\beta|^4 + \mathcal{O}(p^2), \\
F^{\rm QP}_A &= \langle \psi_A(t_k) | \rho^{\rm QP}_A(t_{k+1}) | \psi_A(t_k) \rangle \nonumber \\
&= 1 + (1 - 2|\alpha|^2)p + \mathcal{O}(p^2).
\end{align}

Upon averaging over the Haar distribution with $\langle |\alpha|^2 \rangle = 1/2$ and $\langle |\beta|^4 \rangle = 1/3$~\cite{zyczkowski2005}, the fidelities become:
\begin{align}
\langle F_A \rangle &= 1 - \frac{4}{3}p + \mathcal{O}(p^2), \\
\langle F^{\rm QP}_A \rangle &= 1 + \mathcal{O}(p^2).
\end{align}

In conclusion, the MAQCY protocol, through active error mitigation using erasure and auxiliary storage, achieves superior fidelity scaling$-$preserving quantum information to $\mathcal{O}(p^2)$$-$compared to unprotected single-atom systems.

\end{document}